\documentclass[12pt]{elsarticle}

\usepackage[english]{babel}
\usepackage[utf8]{inputenc}
\usepackage{amsmath}
\usepackage{natbib}
\usepackage{float}
\usepackage{amssymb}
\usepackage{graphicx}
\usepackage{url}
\usepackage{caption}
\usepackage{diagbox}
\usepackage{setspace}
\usepackage{hyperref}
\usepackage{url}

\usepackage{ulem}

\begin{document}

\def\spacingset#1{\renewcommand{\baselinestretch}%
{#1}\small\normalsize} \spacingset{1}

  \title{\bf Using Localized Twitter Activity for Red Tide Impact Assessment}
    \author[rvt]{A. Skripnikov\corref{cor1},\fnref{fn1}}
\author[rvt]{N. Wagner}
\author[rvt1]{J. Shafer}
\author[rvt2]{M. Beck}
\author[rvt2]{E. Sherwood}
\author[rvt2]{M. Burke}

    \cortext[cor1]{Corresponding author.}

\fntext[fn1]{Postal address: New College of Florida, Heiser Natural Sciences Complex, Room E156, 500 College Dr, Sarasota, FL 34243. \ \ Email: askripnikov@ncf.edu. \ \ Tel. : 352-283-2960}

\address[rvt]{
New College of Florida, Division of Natural Sciences, 500 College Dr, Sarasota, FL 34243, USA.
} 

\address[rvt1]{
Science and Environment Council of Southwest Florida, 1530 Dolphin Street, Suite 4, Sarasota, FL 34236, USA.  
}

\address[rvt2]{
Tampa Bay Estuary Program, 263 13th Ave S, St. Petersburg, FL 33701, USA.
}


\bigskip
\begin{abstract}
Red tide blooms of the dinoflagellate {\it Karenia brevis (K. brevis)} produce toxic coastal conditions that can impact marine organisms and human health, while also affecting local economies. During the extreme Florida red tide event of 2017-2019, residents and visitors turned to social media platforms to both receive disaster-related information and communicate their own sentiments and experiences. This was the first major red tide event since the ubiquitous use of social media, thus providing unique crowd-sourced reporting of red tide impacts. We evaluated the spatial and temporal accuracy of red tide topic activity on Twitter, taking tweet sentiments and user types (e.g. media, citizens) into consideration, and compared tweet activity with reported red tide conditions, such as {\it K. brevis} cell counts, levels of dead fish and respiratory irritation on local beaches. The analysis was done on multiple levels with respect to both locality (e.g. entire Gulf coast, county-level, city-level, zip code tabulation areas) and temporal frequencies (e.g. daily, every three days, weekly), resulting in strong correlations between local per-capita Twitter activity and the actual red tide conditions observed in the area. Moreover, an association was observed between proximity to the affected coastal areas and per-capita counts for relevant tweets. Results show that Twitter is a reliable proxy of the red tide’s local impacts and development over time, which can potentially be used as one of the tools for more efficient assessment and a more coordinated response to the disaster in real time.
\end{abstract}


\maketitle




\section{INTRODUCTION}
\label{sec:Introduction}

Red tide algal blooms associated with marine dinoflagellates are regular occurrences in coastal environments where oceanographic conditions, nutrient availability, and planktonic community interactions create conditions favorable for growth.  Increases in the frequency, duration and severity of these events have occurred over the years (\citeauthor{tester1997gymnodinium}, \citeyear{tester1997gymnodinium}, \citeauthor{kirkpatrick2004literature}, \citeyear{kirkpatrick2004literature}, \citeauthor{glibert2005global}, \citeyear{glibert2005global}), stimulated by multiple factors related to coastal land development and climate change (\citeauthor{glibert2020harmful}, \citeyear{glibert2020harmful}).  Toxins (\citeauthor{lin1981isolation}, \citeyear{lin1981isolation}, \citeauthor{poli1986brevetoxins}, \citeyear{poli1986brevetoxins}) produced by these species or degradation of water quality conditions with cell
decomposition can have multiple negative impacts on marine environments and human communities in coastal areas. Large die-off events of marine organisms are commonly observed
with red tides (\citeauthor{sievers1969comparative}, \citeyear{sievers1969comparative}, \citeauthor{steidinger1996dinoflagellates}, \citeyear{steidinger1996dinoflagellates}),  including fish, marine associated birds, turtles, and large mammals (e.g., cetaceans, sirenians).  Bioaccumulation of neurotoxins can
further alter ecosystem trophic dynamics (\citeauthor{pierce2001innovative}, \citeyear{pierce2001innovative}).  Coastal conditions
for humans during ride tides are also negatively impacted, as toxins are aerosolized out of the water column contributing to respiratory or skin irritation that can lead to hospitalizations (\citeauthor{backer2005occupational}, \citeyear{backer2005occupational}, \citeauthor{fleming2005initial}, \citeyear{fleming2005initial}, \citeauthor{fleming2007aerosolized}, \citeyear{fleming2007aerosolized}, \citeauthor{milian2007reported}, \citeyear{milian2007reported}).  Economic impacts are also observed, as tourism or waterfront business revenues are reduced in areas affected by red tide.

Red tides from {\it K. brevis} have been observed on the southwest coast of Florida for over a century. Similar to global trends in harmful algal bloom (HAB) occurrences, these events have increased in severity, frequency, and geographic extent in recent years.  While blooms originate offshore on the West Florida Shelf, onshore winds and currents can bring red tide blooms inshore to beaches and estuaries along Florida’s Gulf Coast, where they may be exacerbated by coastal nutrient sources.  Extensive hydrologic modification of southwest Florida has altered the quantity, location, and timing of nutrient delivery from background conditions. Land use conversion of wetlands to agriculture and urbanization of coastal areas has further altered drainage patterns. The impacts of these changes on coastal water quality in southwest Florida have  been exacerbated by  climate change, as seasonal  temperature and precipitation patterns have deviated from historical averages. Although no single cause
has been identified, the synergistic effects of multiple stressors have likely contributed to significant red tide events.  One of the longest and most severe events to date persisted on the coast of southwest Florida for 16 months from Fall of 2017 to early 2019.

Given the potential for large negative effects of red tide events on the environment, economy and public health, additional datasets that can augment conventional monitoring data could improve how local governments plan and respond to these events. Local governments and regional planning councils rely on regularly updated and routine monitoring data to respond appropriately to environmental disasters. These data inform how future response is managed to anticipated events, as well as how real-time response to ongoing events changes relative to where impacts are most observed. For red tide events, local governments may require data on where blooms are most severe to inform clean-up events (e.g., in response to fish die-offs) or to issue public safety announcements regarding beach closures or air quality advisories.  Real-time rapid assessment of in situ conditions can inform these decisions, whereas more detailed information about oceanographic conditions and phytoplankton cell counts that require lab processing can be used for forecasting where bloom events are likely to occur in the future (\cite{GomxHABForecast}, \cite{USFRedTideTracking}). Moreover, conducting keyword analysis of social media
messages could help in determining the issues of public’s utmost concern (e.g. is it economic impacts? health consequences? general concern for the environment?), hence informing decisions on the optimal distribution of the resources and funding (e.g. assisting waterfront
businesses, distributing protective equipment, organizing marine debris cleanup, etc).

The recent bloom in southwest Florida was the first major red tide event since broad public use of social media platforms, offering potentially unique opportunities to assess complementary sources of information that can aid management response to natural disaster events.  During natural catastrophes like floods, earthquakes, and hurricanes people have been turning to platforms such as Twitter
 (\citeauthor{eismann2016collective}, \citeyear{eismann2016collective},  \citeauthor{kryvasheyeu2016rapid}, \citeyear{kryvasheyeu2016rapid},
\citeauthor{mukkamala2017social}, \citeyear{mukkamala2017social},  \citeauthor{zou2019social}, \citeyear{zou2019social}), Facebook (\citeauthor{bird2012flooding}, \citeyear{bird2012flooding}, \citeauthor{eismann2016collective}, \citeyear{eismann2016collective}, \citeauthor{bhuvana2019facebook}, \citeyear{bhuvana2019facebook},  \citeauthor{jayasekara2019role}, \citeyear{jayasekara2019role}), Instagram (\citeauthor{sherchan2017harnessing}, \citeyear{sherchan2017harnessing}), Flickr (\citeauthor{liu2008search}, \citeyear{liu2008search},
\citeauthor{preis2013quantifying}, \citeyear{preis2013quantifying}, \citeauthor{chien2017does}, \citeyear{chien2017does}) to share their sentiments and
experiences, making social media a valuable source of information.  There’s been a variety of aspects studied in cases of utilizing social media within the context of disaster events, ranging from collective behavior analysis
 (\citeauthor{liu2008search}, \citeyear{liu2008search}, \citeauthor{eismann2016collective}, \citeyear{eismann2016collective}), to risk and crisis communication (\citeauthor{veil2011work}, \citeyear{veil2011work}, \citeauthor{bird2012flooding}, \citeyear{bird2012flooding}), to relief effort coordination (\citeauthor{gao2011harnessing}, \citeyear{gao2011harnessing}, \citeauthor{purohit2014emergency}, \citeyear{purohit2014emergency}, \citeauthor{bhuvana2019facebook}, \citeyear{bhuvana2019facebook}), among others. Meanwhile, in this work we intend to focus on studying the spatio-temporal correlations between localized social media activity and disaster impacts to respective areas
 (\citeauthor{preis2013quantifying}, \citeyear{preis2013quantifying},
\citeauthor{kryvasheyeu2016rapid}, \citeyear{kryvasheyeu2016rapid},
\citeauthor{chien2017does}, \citeyear{chien2017does},
\citeauthor{chen2020assessing}, \citeyear{chen2020assessing}), along with conducting keyword and sentiment analysis for gauging public’s situational awareness during the disaster (\citeauthor{verma2011natural}, \citeyear{verma2011natural}, \citeauthor{yin2012esa}, \citeyear{yin2012esa}, \citeauthor{mukkamala2017social}, \citeyear{mukkamala2017social}). That would serve to provide an additional use case and partially confirm some of the research done in \citeauthor{preis2013quantifying}, \citeyear{preis2013quantifying},
\citeauthor{kryvasheyeu2016rapid}, \citeyear{kryvasheyeu2016rapid}, when user activity on Twitter and Flickr was shown to be a good proxy for evaluating local damages during hurricane Sandy, or the study conducted in \citeauthor{chen2020assessing}, \citeyear{chen2020assessing}, where tweets containing certain keywords were shown to correlate with road closures due to hurricane Harvey flooding.

Our study focused on Twitter - a “microblogging” service used by over 330 million people worldwide to share short strings of text or “tweets” to convey opinions or ideas on any
topic.  Besides its efficiency of communication (due to character restrictions imposed on tweet content), a key component of Twitter is the ability to share or “retweet” information from other users such that a single message can reach a much wider audience than the user’s immediate followers.  Tweets can also be used to assess a response in time (i.e., how do people react after an event) that can be traced to a geographic location if, for example, a tweet  is spatially referenced by  the user.  Thus,  interpretation of tweet  content across multiple users, relative numbers of retweets, spatial, and temporal information can provide
an assessment of topic importance that can guide response and recovery efforts. The use of Twitter to both gauge the impacts and inform disaster response has received some attention in the recent literature.  For example, links between Twitter activity and localized storm damages (\citeauthor{kryvasheyeu2016rapid}, \citeyear{kryvasheyeu2016rapid}) and community response to storm events when power
outages prevented use of conventional communication outlets (\citeauthor{pourebrahim2019understanding}, \citeyear{pourebrahim2019understanding}) have both been demonstrated with an analysis of tweet data during hurricane Sandy. While storm events unfold over days, red tide events can last several months and thus introduce a significant temporal as well as spatial component to the impacts.

Here, we present a multi-level analysis of Twitter activity as it relates to the Florida 2017-2019 red tide event. Our overall goal was to evaluate the use of Twitter as a potential complement for data on real-time conditions as measured by in situ observations (e.g., phytoplankton cell counts) or reported conditions relevant to public health (e.g., beach conditions). The intent is not to supplant existing methods for tracking status and trends of red
tide conditions, but to understand social responses to these events that may have management implications for improving communication about red tide, both general education and
real time conditions, and for prioritizing prevention or mitigation actions by  local governments.  The use of Twitter as a proxy for real time conditions could also augment existing datasets by providing either confirmatory, or in some cases anticipatory, information in the absence of in situ data. For example, this may have value if regular monitoring programs for tracking conditions do not exist or lab processing times prevent rapid response to current
conditions. Lastly, we analyzed sentiment and topical nature of the tweets for the purposes of gauging public’s situational awareness during the event, determining the issues of concern (environment, health,  economy,  government), and potentially improving the accuracy of
Twitter metrics as proxies for local red tide conditions.

\section{MATERIALS AND METHODS}
\label{sec:MaterialsMethods}

\subsection{Red tide local conditions data.}

For local beach conditions data \citep{LocalBeachData}, such as dead fish and respiratory irritation levels, we used daily reports from 12 main beaches, which included six in Sarasota county (Manasota, Venice, Venice/North Jetty, Nokomis, Siesta Key, Lido Key beaches), two in Manatee
(Coquina, Manatee beaches), and four in Pinellas (Pass-a-Grille, Treasure Island, St Pete, Clearwater beaches). When accumulating the data by counties, the observed dead fish and respiratory irritation levels were averaged across the beach pertaining to one county.  On
city and zip code tabulation area (ZCTA) levels, each city and ZCTA were either assigned the average reported conditions from the beaches nearby, or assigned red tide impact of zero if they weren’t near any of those beaches.

{\it K. brevis} cell count data \citep{KBrevisData} was not as consistently available as the local beach conditions
data, taking on the form of samples collected sporadically all across the southwest Florida coast, with the exact longitude and longitude coordinates of sample locations provided (see Figure \ref{fig:FigureSupp2}(A) in the Appendix).	We managed to define polygons on the map that allowed to assign all {\it K. brevis} samples taken with that specific area to the nearest county, but doing it on city- and ZCTA-levels was impractical, hence we stuck to county-level data for {\it K. brevis}. Moreover, when summarizing {\it K. brevis} cell counts over a week or a three-day period, we had to account for the fact that some counties had more samples taken (e.g. Sarasota), and most samples generally showed 0-cell counts. It meant that,
if we were to simply take the mean cell count across all samples, counties with larger number of samples could have their means severely down-graded due to a huge number of 0-cell count samples, despite having several other areas of really high cell counts. Instead, we evened the playing field by only considering the five largest cell counts during that period in that particular county.

\subsection{Raw Twitter data collection.}

The raw data for the 2017-2019 Florida red tide event was obtained via Twitter’s Full Archive Premium search. First and foremost, to guarantee that the tweet is pertinent to
the red tide event, we searched for those that had either ”red tide” or ”redtide” mentions in their content.  Secondly, given that the focus of the study was on evaluating local impacts of the red tide on the five counties situated on Florida’s Gulf coast surrounding Tampa
Bay and Sarasota Bay (Pasco, Hillsborough, Pinellas, Manatee and Sarasota), we needed tweets that also had an affiliation with that particular area. It was accomplished by running
search queries that match tweets to a location in two different ways. First is called a ”place” match, which pertains to tweets that were explicitly geo-tagged by the user as being sent from the location of interest.  Second is a ”geoprofile” match, corresponding to tweets from users that had a relevant location specified in their profile description.  Geoprofile matches represent implicit association with the area of interest because, unlike explicit geo-tags, they don’t guarantee that the tweet was sent from the location specified in the user’s profile
description. Nonetheless, due to scarcity of place matches (only about 1.5\% of tweets are explicitly geo-tagged), geoprofile matches served to both considerably augment the amount of potentially pertinent tweets, and to construct several alternative Twitter metrics to be compared in terms of reflecting the actual red tide impacts.	Besides different location-matching criteria, tweets were also subdivided by user account types (e.g.  media, citizen, etc) to see if Twitter activity by specific subgroups of users serves as a more reliable proxy of the conditions. For the user privacy protection purposes we won’t be releasing the exact criteria of determining account type assignment, bar the fact that the account verification status was used to distinguish most citizen accounts (unverified status) from those of public interest, e.g. media, political figures, government agencies, visitor bureaus, that get the official verification stamp (blue tick mark, signifying verified status) from Twitter. Lastly, we only included tweets posted between May 15 of 2018 and May 15 of 2019, giving a sufficient window to witness both the anticipation period (preceding the first coastal evidence of red tide in June 2018) and the potential long-term after-effects (several months after the last observation of red tide in January 2019 near Sarasota area) for the time window when red tide was most active.

\subsection{Twitter data cleaning.} 

To verify relevance of all the matched tweets, post-collection data cleaning was performed. Presenting the biggest issue were the tweets containing phrases like ”Red Tide Rick” and ”Red Tide Party”, referencing nicknames for Florida’s then-governor Rick Scott and the Republican Party, respectively. Given that 2018 Florida red tide event happened to strongly overlap with the local elections, such references could be found to match better with the dates of political events (e.g. general election or campaign events) rather than the actual local red tide impacts (see supplementary materials). To avoid conflation of public’s concern and reports about local conditions with someone using it as a political statement, we simply disposed of tweets that, outside of political nicknames, contained no other mention of the actual red tide event.

Having cleaned the overall tweet corpus, the loads that were obtained via different location-matching criteria - explicit geo-tags and geoprofile matches - were merged to avoid
double-counting the tweets matched by more than one criteria.  In the meantime, if we encounter tweets that happened to be matched to one area via explicit geo-tag and to a different area via geoprofile (e.g. a tweet geo-tagged in Bradenton, while user’s profile location says ”Sarasota”), we prioritize the place match in determining the ultimate location affiliation of the tweet.

Another round of cleaning had to do with tweets being assigned as a geoprofile match to Hillsborough county purely based on the user specifying ”Tampa Bay” as their profile
location, which could have easily pertained to Pinellas county as well.  Therefore, we reassign these tweets as geoprofile matches to the greater Tampa Bay area, with the credit subsequently shared between Hillsborough and Pinellas (see the details of credit-sharing scheme in Section \ref{sec:CreditSharing}).

Having completed all the cleaning stages, we ended up with a grand total of 18082 tweets geo-matched to our five counties of interest. Among these, 1295 were explicit geo-tags, while
the other 16787 were matched solely based on geoprofile location information.  This tweet load included original tweets (8841, $49\%$ of the load), replies (1784, $10\%$) and retweets (7457, $41\%$).

\subsection{Per-capita tweet counts and credit sharing among several counties.}
\label{sec:CreditSharing}

Among other considerations, because the number of Twitter users can vary greatly depending on the location,  we  considered metrics on per-capita basis,  normalizing  by  the population of that respective area.  Tables 2-4 in the Appendix point to per-capita local tweet counts presenting a much more reliable picture of the actual red tide conditions in the area. Data from US Census Bureau website \citep{USCensusBureau} was used to obtain population estimates for our five counties of interest in 2018.  While we are aware that Twitter only allows users of age 13 and older, the exact percentages of 13+ population weren’t provided, and we simply decided to go with the overall population as a sound proxy. Hillsborough county had nearly
1.5 million people (including the highly populated Tampa metro area), Pinellas being second with 973 thousand, Pasco third with 539 thousand, Sarasota - 426 thousand, and Manatee county was last with about 400 thousand.

Credit-sharing for tweets matched to the general Tampa Bay area took place commensurate to the relative twitter-eligible population sizes of Hillsborough (60\%) and Pinellas (40\%) counties. So, if we had a 100 tweets matched to Tampa Bay area, 60 of them are credited to Hillsborough, with the remaining 40 - to Pinellas.

In the meantime, when working on city- and ZCTA-level data, we’ve also used US Census Bureau data with 2018 population estimates for all the cities in the Tampa Bay
surrounding area. Moreover, we made sure to ”smooth” the populations of smaller cities to be considered as a part of the metro area around the closest large city, avoiding for such tiny towns as Bradenton Beach or Anna Maria (with populations of under 1000) to be blown out of proportion as a result of per-capita calculations.  Hence, all cities belonging to the same metro area were divided by the population of that metro area (e.g. in case of Bradenton
Beach, we divided its counts by the total cumulative populations of all surrounding cities, such as Bradenton, West Bradenton, Palma Sola, Bradenton Beach itself, etc), while ZCTA counts were divided by the population of the metro area that it was associated with.

\subsection{Sentiment analysis.} 
\label{sec:SentAnalysis}

The main goal of sentiment analysis is to quantify the emotion expressed within the text. Depending on whether the message is positive, negative or neutral in nature, an appropriate numerical value is assigned to reflect that sentiment. In the case of our data, monitoring temporal dynamics of the sentiment for red tide related tweets in a given location might aid in determining whether the area is currently being affected (”Siesta Key beach is littered with
dead fish, smell is horrible \#redtide”), recovered from it (”Red tide is gone from Madeira Beach”), or avoiding it altogether (”No red tide in Clearwater!”).  In particular, we will be looking at correlations between the sentiment expressed by Twitter users and the actual red tide conditions observed in the area at the time.

To assign sentiment values, we use the methodology implemented in $sentimentr$ package \citep{sentimentr} from $R$ Statistical Software \citep{RCoreTeam}. It relies on a combination of pre-defined polarized word lexicon,
vocabulary of valence-shifters, and question weight. Polarized words (or phrases) are assigned a
positive or negative value, with magnitude representing the strength of expressed emotion (e.g.
”bad” is $-0.75$, while ”disgusting” is $-1.0$; ”good” is $0.75$, while ”gorgeous” is 1). Valence-shifters
are words (or phrases) that might change either the direction or strength of the polarized  words
found in their vicinity.  That includes complete negation (e.g.  ”not”, ”isn't”, ”hasn't” would
flip the sign of polarized words nearby), amplification/de-amplification (e.g. ”very”, ”strongly”,
”completely” would increase the magnitude of polarized words nearby, while ”hardly”, ”barely”
- decrease it), adversative conjunction (e.g. ”but”, ”however”, ”although” would increase the
strength of the following polarized words, while decreasing it for the preceding ones). Moreover,
the questions can be downweighted, due to expressing less conviction as opposed to declarative
statements (e.g. ”is red tide really bad there?” as opposed to ”red tide is really bad there.”). In
our case, we set question weight at $0.25$, meaning that the total sentiment score of a question
would be divided by 4. Lastly, we also treated occurrences of multiple periods in a row (e.g.
”this red tide...”, ”water is discolored...”) as conveying a negative emotion (generally used to
indicate exasperation, annoyance, frustration), so we decreased the sentiment value by $0.15$ per
each such occurrence in a tweet.

In case of the red tide event, we needed to customize the polarized word lexicon for it to be more applicable. As an example, we needed to dispose of the word ”bloom” as a positive, given that in our context it pertains to algae bloom - an occurrence that’s negative in relation to HAB events.  We likewise needed to associate words like ”impact” and ”affect” with the negative effects the red tide had on the local environment, economy and health. Moreover,
occurrence of phrases such as ”low/medium/high levels”, ”low/medium/high concentrations” exclusively pertained to the presence of {\it K. brevis} cell counts, another negative event.  On the other hand, all phrases pointing to the absence of red tide were given a strong positive
score (”no more red tide”, ”red tide is gone”, ”no signs of red tide” all at +1).

We couldn’t possibly verify the assigned sentiment scores for every single tweet (total of over 10 thousand tweets with distinct contents), but we randomly sampled a 1000 tweets (practice similar to
 \citeauthor{verma2011natural}, \citeyear{verma2011natural}, \citeauthor{mukkamala2017social}, \citeyear{mukkamala2017social})
and meticulously went through them in the best attempt to put together the most appropriate vocabulary of polarized words, valence-shifters and their respective assigned scores.

 \subsection{Relationship between Twitter activity and distance to impacted areas} 
 
 Our analysis of Twitter activity and proximity to the impacted areas relied on geo-coordinates provided by tweet's spatial information (be it that of an explicit geo-tag, or of user profile location) and that of {\it K. brevis} samples. To define areas of high impact, we isolated the weeks containing high {\it K. brevis} cell count ($>1,000,000$ cells/l) for at least one sampling location, and computed minimum geodesic distances between tweet locations and closest site of high {\it K. brevis} levels. To obtain tweet counts (or retweet proportions) for each relevant week, we simply calculated totals of tweets (proportion of retweets, respectively) associated with a unique Twitter-assigned pair of coordinates. The plots observed on Figure \ref{fig: Figure4} were obtained by accumulating these Twitter metrics along with distances across all weeks of observing high {\it K. brevis} cell counts. Analogous work was done when qualifying impacted areas as locations with heavy dead fish levels and intense respiratory irritation, yielding similar results.

  \section{RESULTS.}

\subsection{Cumulative Twitter activity on "red tide" topic across metro areas.}
\label{sec: CumulTwitterActivity}

\begin{figure}[]


\begin{minipage}{1\textwidth}
\ \ \ \ \ \  \ \ \ \ \ \ \ \textbf{A} 

\ \ \ \ \ \  \ \ \ \ \ \ \ \includegraphics[scale=0.7]{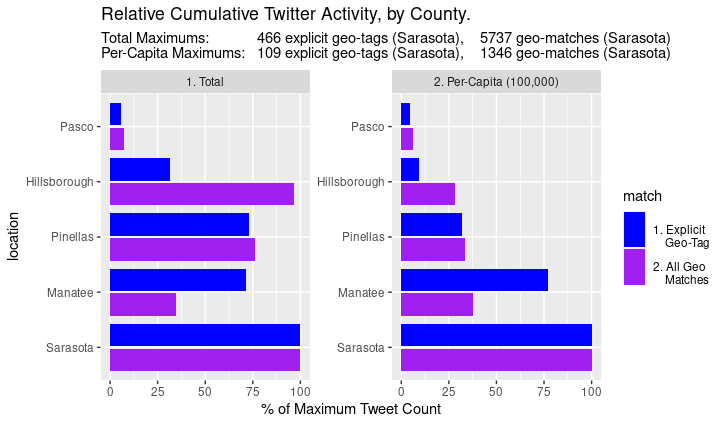} \ \ 


\

\ \ \ \ \ \  \ \ \ \ \ \ \ \textbf{B}

\ \ \ \ \ \  \ \ \ \ \ \ \ \ \ \ \ \ \ \ \ \ \ \ \ \includegraphics[scale=0.7]{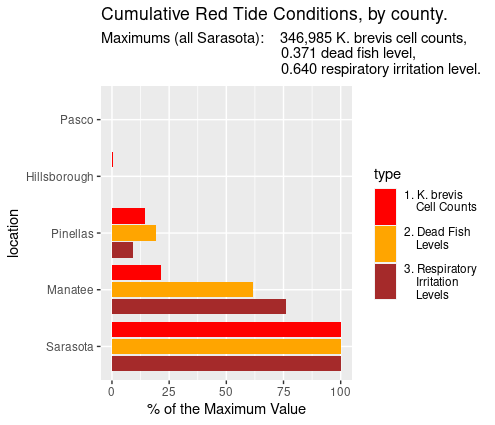}
\end{minipage}

\caption{\textbf{Relative tweet counts and actual red tide conditions, accumulated by each county during Florida red tide 2018 event}. \textbf{A}: Gross (left) and per-capita (per 100,000 population, right) totals, for explicitly geo-tagged tweets only (blue) and all geo-matched tweets (including geoprofile matches as well). \textbf{B}: Beach condition index for dead fish (orange) ranged from 0, ”None”, up to 2, ”Heavy”; beach condition index for respiratory irritation (brown) ranged from 0, ”None”, up to 3, ”Intense”.}
\label{fig: Figure1}
\end{figure}

One of the main goals was to study localized Twitter activity in five Florida Gulf coast counties located in the Tampa Bay surrounding area: Pasco, Hillsborough (including Tampa metro), Pinellas (Clearwater and St Petersburg metro areas), Manatee (including Bradenton metro), and Sarasota (including cities of Sarasota, Venice and Englewood). First, evaluating each area’s total Twitter activity - throughout the entire duration of the Florida red tide event, including the anticipation and after-effects - may be of interest to identify potential correspondence to the actual observed local red tide impacts, such as {\it K. brevis} cell counts, dead fish and respiratory irritation levels. The tweet counts and red tide conditions for these counties relative to their location along the coast can be observed on Figure \ref{fig: Figure1}.

A clear pattern of diminishing red tide impact moving northward from Sarasota to Pasco is reflected on Figure \ref{fig: Figure1}(B) in decreasing number of {\it K. brevis} cell counts, dead fish and
respiratory irritation levels observed in the respective areas.  In the south, Sarasota and Bradenton were most impacted, whereas impacts in Hillsborough and Pasco counties to the north were not as severe.  In contrast, if one were to rely on total tweet counts as the sole proxy for red tide impacts for each metro area - ”Total” plot on Figure \ref{fig: Figure1}(A) -  Hillsborough would stand out as the most active region in terms of Twitter activity, despite receiving minimal effects of red tide. That happens due to this region, that includes Tampa, having by far the largest population, and we alleviate this effect by employing per-capita tweet counts. Results can be observed in ”Per-Capita” Twitter activity plot on Figure \ref{fig: Figure1}(A). We witness a truer reflection of the actual red tide impact ranking for the five counties of interest, with Sarasota showing the highest per-capita tweet counts, while Hillsborough is ranked on par with Pasco with the lowest counts using this adjustment. The explicit geo-tag counts (in blue, Figure \ref{fig: Figure1}(A), ”Per-Capita”), which serve as a more reliable indicator of a
person tweeting from the location of interest, perfectly mirror the county rankings by actual red tide impact.	Unlike all geo-matched tweets (in purple, Figure \ref{fig: Figure1}(A), ”Per-Capita”),
considering explicit geo-tags only leads to Manatee clearly placing as the second most active county on Twitter, while greatly reducing Hillsborough’s Twitter activity, which reflects the severity of actual local conditions reflected on Figure \ref{fig: Figure1}(B). Lastly, although explicitly geo-tagged tweets are rare, they are much faster and easier to obtain in real time, due to Twitter’s Full Premium Archive service allowing for a simpler set of filtering rules for obtaining  relevant geo-tags as opposed to filtering based on user’s geoprofile information.

\subsection{Weekly local Twitter dynamics compared to weekly local red tide conditions.}
\label{sec:WeeklyTBarea}

\begin{figure}[]
\begin{minipage}{0.5\textwidth}

\textbf{A}

\includegraphics[scale=0.63]{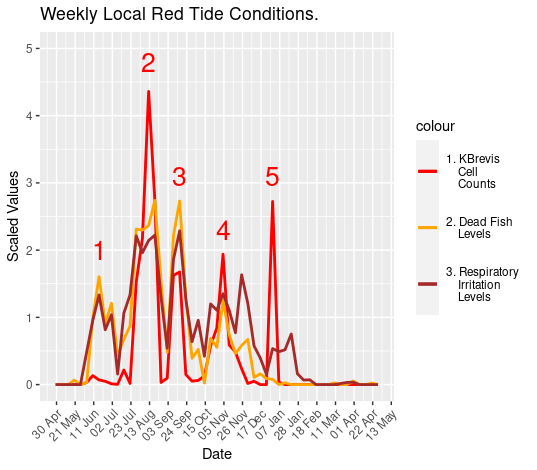} 

\

\textbf{B}

\includegraphics[scale=0.63]{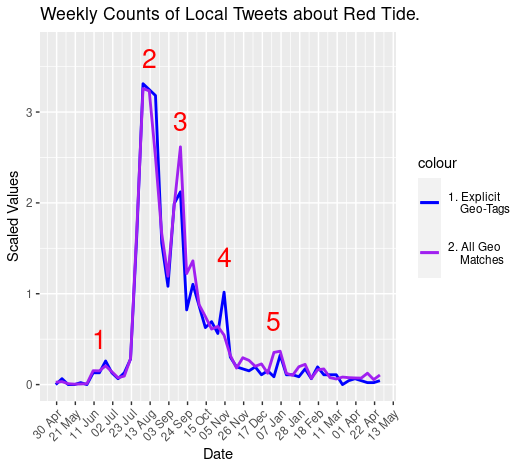}
\end{minipage} \ \ \ \  \ \ 
\begin{minipage}{0.5\textwidth}
\textbf{C}

\includegraphics[scale=0.21]{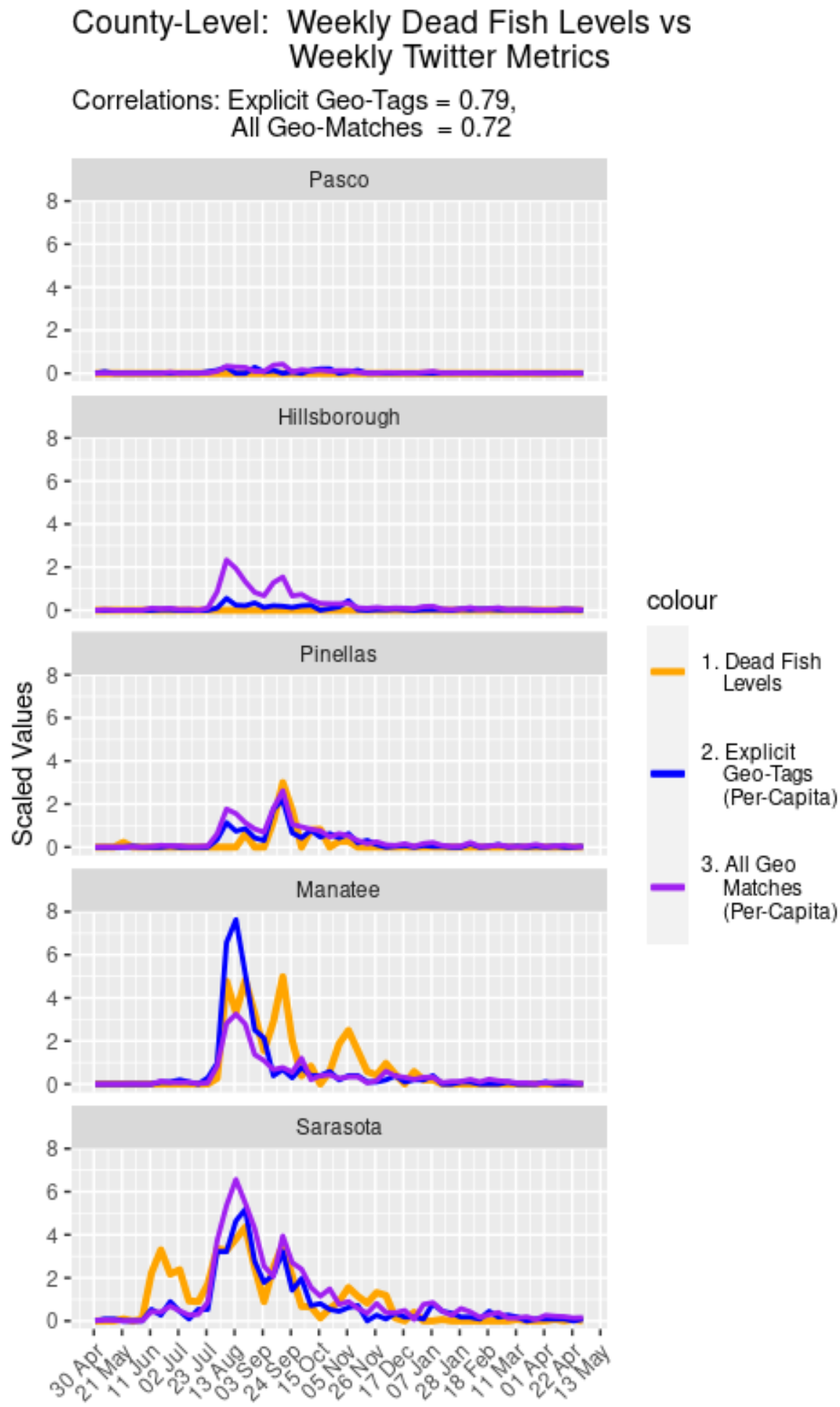}
\end{minipage}

\caption{\textbf{Weekly local red tide conditions and related tweet counts during the 2018 Florida event.} \textbf{A}: Weekly red tide conditions. \textbf{B}: Total tweet counts, for the entire Tampa Bay surrounding area. \textbf{C}: Weekly dead fish levels along with weekly tweet counts, by county.}
\label{fig: Figure2}
\end{figure}

On Figure \ref{fig: Figure2}(A) one may witness time series plots for weekly aggregates of {\it K. brevis} cell counts, dead fish and respiratory levels in the entire Tampa Bay surrounding area from
May 2018 through May 2019.  The five peaks observed correspond to five peak red tide intensity impacts along the studied areas, first one taking place in June (not as strong as others), second - in early August (strongest impact), third - in mid-September, fourth - late October/early November, and fifth - in early January. Figure \ref{fig: Figure2}(B) demonstrates the corresponding weekly Twitter activity during the same period, represented by total counts of tweets that were explicitly geo-tagged in the area (blue), or also matched to it via geoprofile information (purple). One may clearly notice strong temporal correspondence between local Twitter activity and the red tide conditions, matching the second, third and fifth peaks, with a minor bump corresponding to the first peak as well.

Witnessing strong general correlations between Twitter activity and red tide conditions for the entire Tampa Bay surrounding area is promising, but in reality the insights on a more localized scale (i.e.  at the county-, city-level) would be preferable for management and incident response. On Figure \ref{fig: Figure2}(C) we demonstrate the correspondence of our two Twitter metrics - per-capita counts of explicitly geo-tagged and all geo-matched tweets - with the dead fish levels, when broken down by county. In the most affected areas (Sarasota, Manatee and Pinellas), considerable alignment of tweet counts with the periods of high dead fish levels on local beaches is evident, including the minor bump in Sarasota Twitter activity upon the initial sighting of red tide in June.  On the other hand, for the areas like Pasco and Hillsborough - that avoided most of the impact - you might still witness tweets about red tide due to the locals anticipating the red tide reaching their respective area, or simply reporting on what’s happening in the neighboring counties. Nonetheless, it is obvious that explicit geo-tags are less susceptible to that issue compared to all geo-matches (that include the noisy geoprofile-matched tweets), which is strongly exhibited on the example of Hillsborough county, leading to a stronger overall correlation (0.79,  as opposed to 0.72 for all geo-matches).

\subsection{Multi-level spatiotemporal correlations of local Twitter activity with red tide conditions.}

\begin{figure}[]

\textbf{A}

\begin{minipage}{0.47\textwidth}

\includegraphics[scale=0.20]{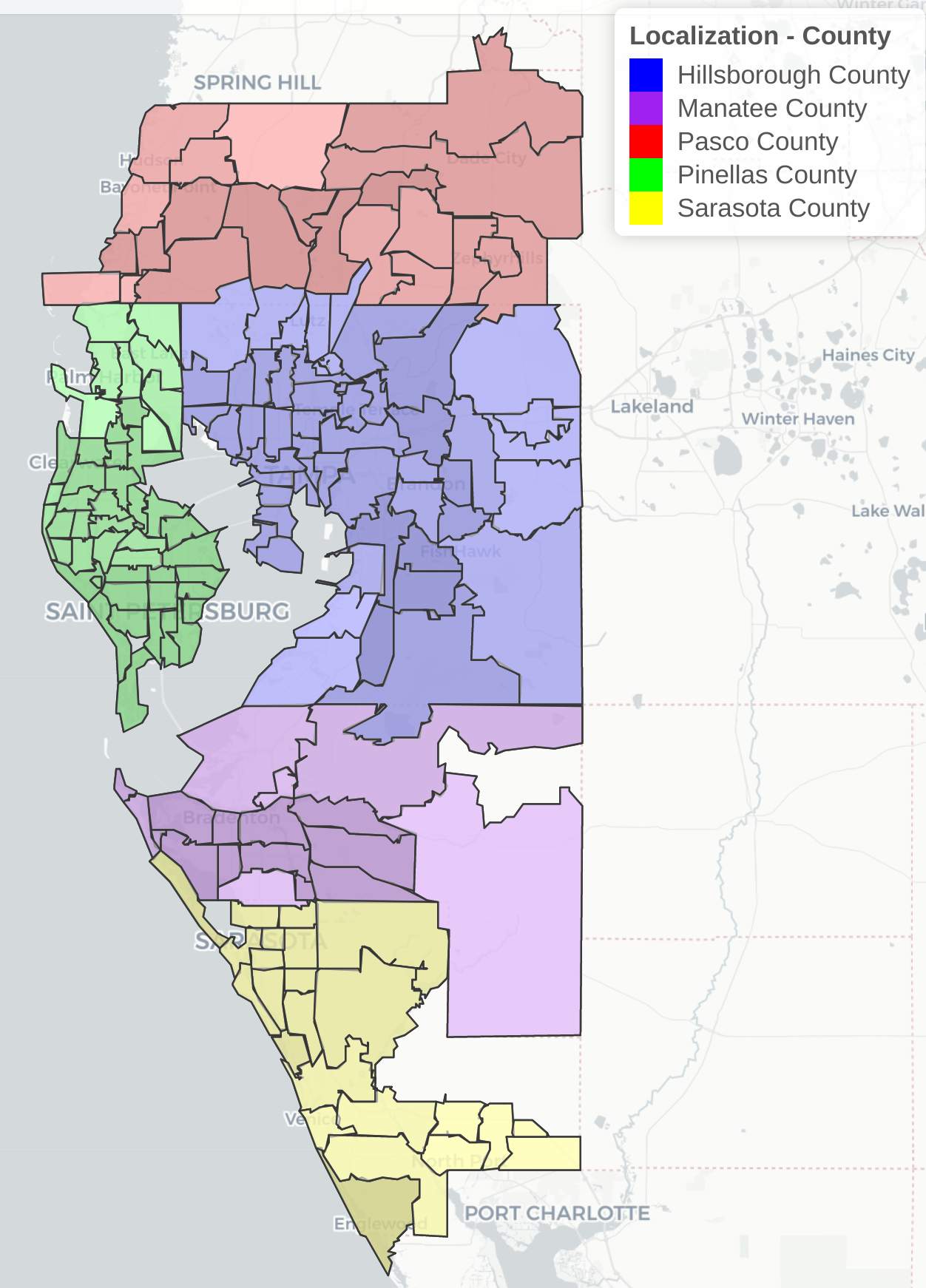}

\end{minipage} \ \ \ \ \ \ \ \ \ \ \ \ \ \ \ \ \ \ 
\begin{minipage}{0.47\textwidth}

\textbf{B}

\includegraphics[scale=0.58]{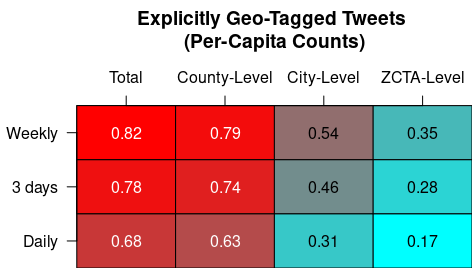}

\

\

\textbf{C}

\includegraphics[scale=0.58]{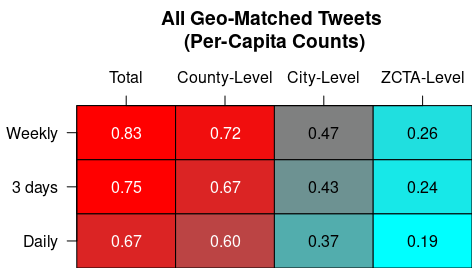}

\end{minipage}
   \caption{\textbf{Multi-level spatiotemporal correlations of local Twitter activity with dead fish levels observed in the area.} \textbf{A}: ZCTA-level boundary map, with all ZCTAs belonging to the same county carrying the same main color. Shade of ZCTA color represents a different primary city this ZCTA is associated with. \textbf{B}: Heatmap-style matrix of spatiotemporal correlations for per-capita counts of tweets explicitly geo-tagged in the area with local red tide conditions, calculated on different levels of locality and time frequency. \textbf{C}: Same as B, but for per-capita counts of all tweets geo-matched to the area, hence also including those matched solely based on the user's geoprofile information.}
   
    \label{fig: Figure3}
\end{figure}

In Section \ref{sec:WeeklyTBarea} we showed what happens if one studies correlations between weekly Twitter activity and local red tide impacts on two localization levels (entire Tampa Bay surrounding area and county-level). Here, we will study the strength of such correlations on several more levels of locality, breaking our data down by 57 primary cities, and subsequently by 166 zip code tabulation areas.  Figure
 \ref{fig: Figure3}(A) shows the aforementioned levels of localization in more detail. Moreover, we experiment with other levels of temporal frequency by looking at the data on a daily basis, or every three days, as opposed to just every week.

On Figures \ref{fig: Figure3}(B-C)  you see the temporal correlations between Twitter activity and dead fish levels observed on local beaches, and how they change depending on the spatiotemporal levels considered.  For the per-capita counts of both the explicitly geo-tagged tweets and all geo-matched tweets, there’s an unsurprising steady decrease in strength of  correlations as we approach more hyper-localized and higher frequency scales.  For example, in case  of
weekly data for explicit geo-tags, we first witness a tiny drop-off in correlations when going from entire Tampa Bay area down to county-level dynamics (0.82 to 0.79), then a much stronger decrease on city-level (0.79 down to 0.54), and further deterioration on ZCTA-level
(0.54 down to 0.35).  Analogous effect is observed when increasing time frequency, e.g. see correlations of explicitly geo-tagged tweets with dead fish levels on county-level go from 0.82 for weekly data, down to 0.78 for three-day totals, to 0.68 for daily data. Nonetheless,
many of these correlations (especially on county-level, regardless of time frequency) are still respectable enough to act as a solid proxy for the actual red tide conditions in respective localities over time, which itself presents an interesting research finding.

Comparing explicitly geo-tagged tweets with all geo-matched tweets in terms of metric robustness across multiple spatiotemporal levels, explicit geo-tags tend to outperform their counterpart virtually all across the board, in addition to being the easier metric to obtain
in real time. Same was shown for respiratory irritation levels, as can be seen on Figure \ref{fig:FigureSupp1} in the Appendix.  Curiously, the only case where all geo-matches were preferred to explicit
geo-tags was county-level correlations of per-capita tweet counts with
 {\it K. brevis} cell counts (see Figure \ref{fig:FigureSupp2} in the Appendix). As described in Section \ref{sec:MaterialsMethods}, {\it K. brevis} samples were collected in a less consistent fashion (arbitrary locations, unstable sampling frequency) than the data on dead fish and respiratory irritation levels (robust daily reports from specific beaches), likely making results for the latter - which showed explicit geo-tags as the preferred proxy for local red tide impacts - more trustworthy.

In addition, we have attempted to focus on correlations of red tide conditions with tweets from specific account types (e.g. only media, only citizens), and also tried studying the delayed reactions and anticipation of red tide impacts via lag and lead correlations, respectively (\citeauthor{spriggs1982lead}, \citeyear{spriggs1982lead}, \citeauthor{dajcman2013interdependence}, \citeyear{dajcman2013interdependence}). There was no evidence found of improved correlations in breaking tweets down by account groupings, and it was shown that Twitter reacted to the disaster in neither delayed nor anticipatory fashion.  For  more details on the analysis, see supplementary materials.

\subsection{Local Twitter dynamics based on proximity to impacted areas.} 

\begin{figure}[]

\begin{minipage}{0.50\textwidth}
\textbf{A}

\includegraphics[scale=0.57]{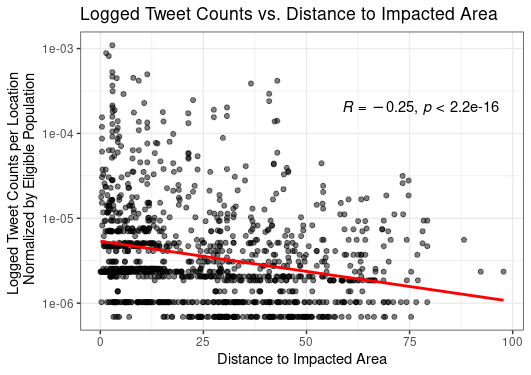}

\

\

\textbf{B}

 \includegraphics[scale=0.57]{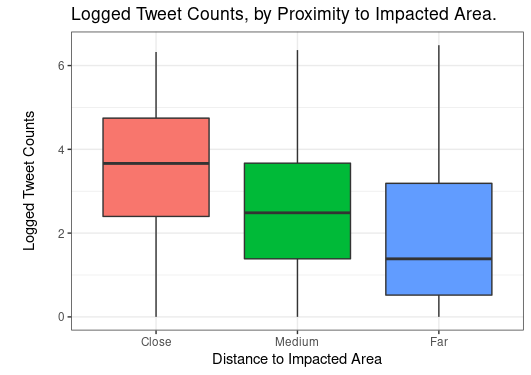}
\end{minipage}  \ \ \  \begin{minipage}{0.50\textwidth}

\textbf{C}

\includegraphics[scale=0.57]{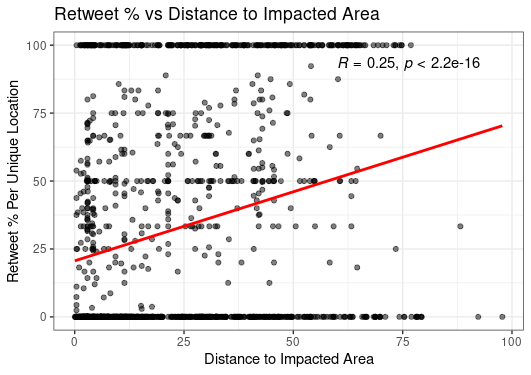}
\end{minipage}

   \caption{\textbf{City-level local Twitter activity and content originality as a function of proximity to the red tide impacted areas}. \textbf{A}: Scatterplot of logged tweet counts against the numerical distance to the currently affected areas. \textbf{B}: Box plots of logged tweet counts coming from locations that are at a close, medium, or far distance from the currently affected areas. \textbf{C}: Proportion of local retweets as a function of proximity to the red tide impacted areas.}
    \label{fig: Figure4}
\end{figure}

One of previous studies on social media use during a disaster event (\citeauthor{kryvasheyeu2016rapid}, \citeyear{kryvasheyeu2016rapid}) specifically investigated the relationship between Twitter activity and proximity to areas of disaster impact. 
We carried out an analogous analysis, measuring distances from tweet locations (cities) to places
experiencing red tide on that given week, and comparing local Twitter activity based on proximity to the disaster. On Figure
 \ref{fig: Figure4} you may witness the logged per-capita counts of local tweets (linear regressions performed on tweet counts as a function of distance warranted a log-transformation) plotted against both the numerical (Figure \ref{fig: Figure4}(A)) and categorical (Close-Medium-Far, Figure \ref{fig: Figure4}(B)) distance to the locations suffering from red tide during a particular week.

In both analyses, we ended up with a statistically significant negative relationship (p-values $< 0.001$ for both), pointing to local Twitter activity decreasing exponentially as we move further away from the disaster areas.  This finding corresponds to the effect observed
 in \citeauthor{kryvasheyeu2016rapid}, \citeyear{kryvasheyeu2016rapid}, where there was an exponential decrease in Twitter activity as one moved further from locations impacted by hurricane Sandy, as long as it was within 300mi of the disaster (beyond that the effect dissipated). While the hurricane Sandy study was on a nationwide scale, including locations removed from disaster impact by thousands of miles, our research focused on a relatively narrow geographical area (only a handful of counties and metros surrounding Tampa Bay). For a smaller scale like ours, categorizing the distance into three bins - close ($<25$mi), medium ($25$-$50$mi), far ($>50$mi) from the red tide - led to a much clearer representation of that negative relationship on Figure \ref{fig: Figure4}(B). In particular, the $95\%$ confidence intervals for pairwise differences in logged tweet counts for Medium-Close ($\in (-1.53, -0.38)$) and Far-Close ($\in (-2.09, -0.93)$) pointed to increasing statistical and practical significance of the difference in tweet counts as we move further away from the red tide-impacted locations.

Another interesting aspect - observed on Figure \ref{fig: Figure4}(C) - was the change in originality of posts as the distance from the red tide increased.  With increasing distance from the impacted areas, an analogous increase in retweet percentage among all red tide-related tweets was observed, meaning that users located further away from the impact were more likely to retweet a post, rather than produce an organic tweet describing their own opinion or experience of the event.  Despite red tides having much longer duration than hurricanes,
this insight was confirmed by work done in \citeauthor{kryvasheyeu2016rapid}, \citeyear{kryvasheyeu2016rapid} on hurricane Sandy, where retweet fraction was shown to clearly increase as location got further removed from the impact. That effect from hurricane Sandy study appeared stronger in big part due to, yet again, the nationwide scale of the respective study, with retweet percentages increasing much faster for locations of $>500$mi away from the impact. Given our focus on a much smaller area, with distances not exceeding 100mi, the magnitude of the observed effect is very comparable
to that of \citeauthor{kryvasheyeu2016rapid}, \citeyear{kryvasheyeu2016rapid} for a similar distance range.  This finding proceeds to reinforce the evidence for potential benefits of accounting for the originality of the content when evaluating local damage and coordinating relief efforts, be it a lengthy or a short-lived disaster event. Granted, to fully confirm this finding, and whether it applies to catastrophes like earthquakes or floods, similar analysis would need to be carried out for other disasters over the last decade.

\subsection{Analyzing sentiments and topical nature of tweets.}

Beyond simply using tweet counts as the way to summarize Twitter activity, we conducted further study into content of the posts, analyzing expressed sentiments and topical nature of the tweets related to red tide.

The goal of sentiment analysis was two-fold: to provide a Twitter metric of total sentiment as an improved proxy for the actual local red tide conditions - with negative values more likely pointing to the red tide affecting the area – and evaluate public’s situational awareness
in regards to the red tide.  To achieve the former, we treated tweets that were reporting on the presence (absence) of the red tide in the area as negative (positive), regardless of whether there was any emotion attached to it (see Section \ref{sec:SentAnalysis} for more details), and subsequently utilized cumulative temporal sentiment scores as the metric to gauge local red tide conditions. On the other hand, to study the level of situational awareness, we conducted keyword analysis to see if the public was acknowledging the event’s magnitude and discussing the main effects and aspects of red tide.

\begin{figure}
    \centering
   \begin{minipage}{0.45\textwidth}
\textbf{A}

\includegraphics[scale=0.2]{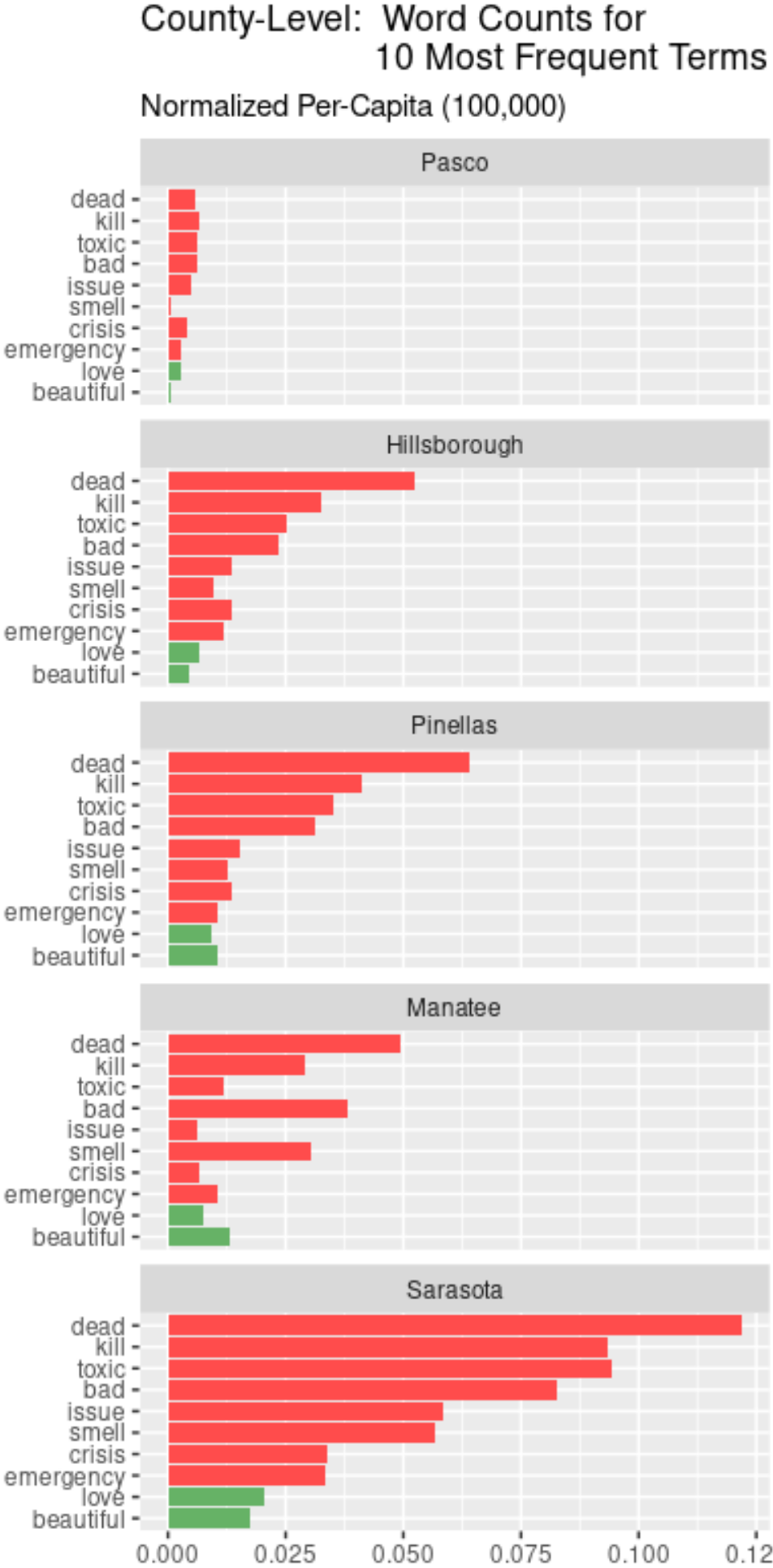}

\

     \end{minipage} \  \ 
     \begin{minipage}{0.45\textwidth}
     
\textbf{B}

\includegraphics[scale=0.209]{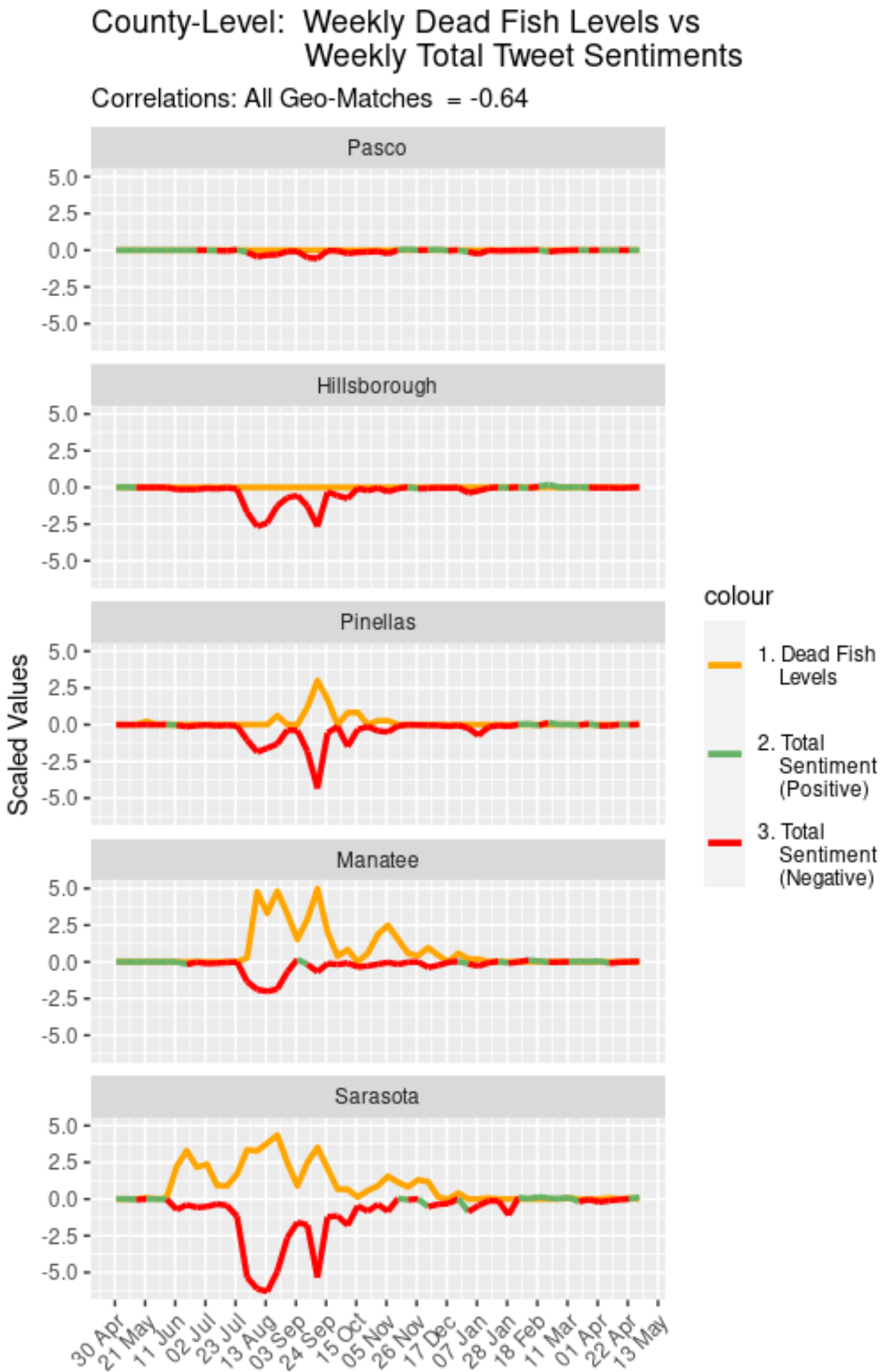}

    \end{minipage}
   
    \caption{\textbf{Twitter sentiment analysis throughout the 2018 Florida red tide event, by county.} \textbf{A}: Word counts of ten most frequent positive and negative terms encountered in tweets associated with respective locations. \textbf{B}: Weekly dead fish levels observed, along with per-capita total weekly sentiment from tweets geo-matched to each county.}
    \label{fig: Figure5}
\end{figure}

On Figure \ref{fig: Figure5}(B) you may witness the correspondence between per-capita weekly total sentiments for geo-matched tweets (mostly in red, due to prevalence of negative sentiment) and dead fish levels for our five counties of interest.  There are several clear instances of high dead fish levels matching with the strong negative sentiment in tweets posted that week, especially for the affected areas of Sarasota, Pinellas and (to lesser extent) Manatee. Additionally, on Figure \ref{fig: Figure5}(A) one may notice the county-level per-capita frequencies of the ten most mentioned positive and negative words. The increase in negative terms is evident when moving north (Pasco) to south (Sarasota), with “dead” and “kill” being the most encountered words, likely due to describing marine life casualties of red tide. Moreover, the word ”smell” shows a strong increase in usage 
when moving towards most affected areas, likely resulting from high dead fish levels in such locations. Lastly, robust usage of terms like “toxic”, “crisis”, “emergency” indicates public’s situational awareness pertaining to both the underlying biological process (toxins killing marine life) and severity of the event (one of the most intense red tides in Florida
history, see \citeauthor{karnauskas2019timeline}, \citeyear{karnauskas2019timeline}). This finding partially updates and augments the study by \citeauthor{kuhar2009public}, \citeyear{kuhar2009public} on the public’s perception of risks related to red tide, where it was concluded that individuals didn’t appear to possess the most updated and adequate information about Florida red tides and their impacts.

\begin{figure}[]
    \centering
    \begin{minipage}{0.47\textwidth}

\textbf{A}

 \includegraphics[scale=0.6]{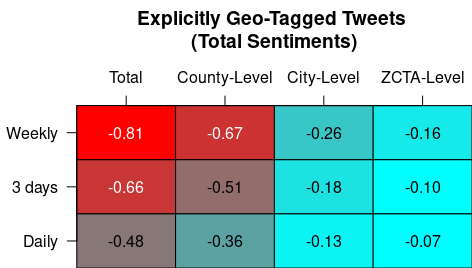}
 \end{minipage} \ \ \ \ \
 \begin{minipage}{0.47\textwidth}
 \textbf{B}
 
 \includegraphics[scale=0.6]{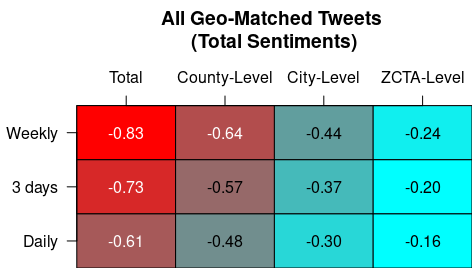}
  \end{minipage}
    \caption{\textbf{Multi-level spatiotemporal correlations of local Twitter sentiments with dead fish levels observed in the area.} \textbf{A}: Heatmap-style matrix of spatiotemporal correlations for per-capita total sentiments of tweets explicitly geo-tagged in the area with local red tide conditions, calculated  on different levels of locality and time frequency. \textbf{B}: Same as A, but for all tweets geo-matched to the area. 
    } 
    \label{fig: Figure6}
\end{figure}

With all of that said, tweet sentiments failed to outperform basic tweet count metrics in regards to correlations with local red tide conditions. For example, looking at the heatmaps of multi-level spatiotemporal correlations for tweet sentiments with dead fish levels on Figure \ref{fig: Figure6}, virtually all correlations were worse (in absolute value) than their tweet count analogs (those shown on Figure
 \ref{fig: Figure3}(B-C)). Add on top the fact that tweet counts are much easier to obtain than sentiment scores, count metrics appear as the best choice for both efficiency and effectiveness from a pure impact assessment standpoint. Regardless, it is still curious to
note that for Twitter sentiment metrics, as opposed to counts, all geo-matched sentiments consistently outperform the explicit geo-tags, by a respectable margin.  This aspect is confirmed on the example of multi-level spatiotemporal correlations for tweet sentiments with
both respiratory irritation levels and {\it K. brevis cell} counts observed locally (Figures \ref{fig:FigureSupp1} and \ref{fig:FigureSupp2} in the Appendix).

In addition to sentiment analysis, we considered categorizing tweet contents by types of concerns that people expressed at the time of the disaster.  Table \ref{tab: Table1} shows red tide  related tweets to contain by far the largest number of unique words and mentions corresponding to environmental concerns (talking about beaches, fish, water, climate, pollution, etc). Health issues (citing respiratory issues, toxicity of red tide, etc) place a distant second in mentions,
and economic concerns (discussing damages to local businesses) third, with both topics being brought up consistently nonetheless. This topical ranking falls in line with the results from \citeauthor{li2015risk}, \citeyear{li2015risk}), where the newspaper coverage of a 2005-2006 {\it K. brevis} bloom was found to predominantly focus on the associated environmental risks (80\% of stories mentioning those), followed by health concerns (48\%), and economic impact was covered the least (30\%). In the meantime, local government was also mentioned consistently, which was likely a function of both the public’s frustration with the politicians, and the Florida 2018 general election, where the latter heavily overlapped with this red tide event (see supplementary materials for further discussion of this point).


\begin{table}[]
    \centering
    \begin{tabular}{llll}
 Topic & Term Examples & \# Unique Terms & \# Mentions \\
\hline
  Environment  & beach, fish, water, climate, pollution  & 48  & 27,430 \\
  Health &  respiratory, toxic, sick, harmful & 34 & 8,267\\
  Economy & business, tourist, restaurant, budget & 25 & 4,970 \\
  Government & governor, policy, vote, election & 43 & 3,357 \\
\end{tabular}
\caption{\textbf{Most cited concerns on Twitter during Florida red tide event of 2017-2019, by category.}}
\label{tab: Table1}
\end{table}

\section{DISCUSSION}

As one of the main conclusions to our study, we discovered that, during a lengthy disaster event such as Florida red tide of 2017-2019, Twitter activity can be used to gauge community perceptions of localized coastal impacts, in both comprehensive and discrete spatiotemporal
scales. One can consider several distinct tweet metrics - counts or sentiments of tweets that were either explicitly geo-tagged or geo-matched to the area in any way - for improved representation of the relative damage to respective areas at each moment in time.

When linking red tide impacts of several metro areas of interest on the Florida Gulf coast to community perceptions, we used {\it K. brevis} cell counts observed in local waters, along with dead fish and respiratory irritation levels reported for local beaches. A clear hierarchy of red tide impacts from spatial standpoint emerged, with southern counties (Sarasota, Manatee, Pinellas) getting most affected, while locations to the north (Hillsborough, Pasco) managed to avoid the disastrous effects.  Per-capita counts of tweets geo-matched to each respective area served as a great proxy for the local damage received, improving on gross tweet totals. Stronger correlations were obtained by only utilizing the explicitly geo-tagged tweets, which are much more reliable than the tweets assigned to a metro area utilizing user geoprofile location information alone (similar observation on lower reliability of geoprofile matches was made in
 \citeauthor{chien2017does}, \citeyear{chien2017does}).

While several studies have been done previously on the use of Twitter to both gauge the impacts and inform disaster response
 (\citeauthor{kryvasheyeu2016rapid}, \citeyear{kryvasheyeu2016rapid}, \citeauthor{pourebrahim2019understanding}, \citeyear{pourebrahim2019understanding}, \citeauthor{chen2020assessing}, \citeyear{chen2020assessing}), most of those pertained to instantaneous or short-lived disasters that typically don’t last longer than a week. Red tide, on the other hand, can last several months at a time, henceforth introducing a considerable temporal component alongside an already existing spatial dimension.  To the best of our knowledge, there has only been one study conducted using Twitter data in regards to a similar disaster event
 -  \citeauthor{mascareno2020twitter}, \citeyear{mascareno2020twitter}, investigated Twitter dynamics throughout the duration of the 2016 red tide event at Chiloe Island, Chile. With that said, their work focused predominantly on studying social interrelationships across multiple user groups and how they changed throughout the disaster event, without putting as much emphasis on correlating localized Twitter activity with the actual red tide impacts observed in the area.

In the specific case of 2017-2019 Florida red tide event, we witnessed algal blooms with varying coastal intensities throughout the duration of roughly eight consecutive months, warranting special attention to be paid to understanding temporal dynamics of red tide development and coastal community responses via Twitter activity over time. When looking at the entire Tampa Bay surrounding area (Figure \ref{fig: Figure2}), strong correlations between weekly per-capita tweet counts and corresponding local conditions were observed, with explicitly
geo-tagged tweets serving as the most reliable proxy, while also presenting the most efficient metric due to simplicity of data collection in real time.  

To draw a comparison with previous studies on social media usage during short-lived disaster events, Florida red tide temporal dynamics are somewhat similar to those of hurricanes, floods or earthquakes solely in the short windows during peak impacts, where spikes in Twitter activity correspond to spikes in respective conditions (dead fish and respiratory irritation levels, {\it K. brevis} cell counts), as was shown on Figure \ref{fig: Figure2}. Nonetheless, there are still
several critical distinctions. First, even the duration of a single spike in red tide conditions is much longer than that of instantaneous disaster impacts. To observe this particular aspect, see daily dynamics of the red tide resurfacing in Sarasota county during September of 2018 on Figure \ref{fig:DailyTemporalDynamics} in the Appendix, and compare these to the timelines of hurricane Sandy (see \citeauthor{preis2013quantifying}, \citeyear{preis2013quantifying}, \citeauthor{kryvasheyeu2016rapid}, \citeyear{kryvasheyeu2016rapid}) or hurricane Harvey flooding (see \citeauthor{zou2019social}, \citeyear{zou2019social},  \citeauthor{chen2020assessing}, \citeyear{chen2020assessing}), where all the impacts took place over less than three-to-four days. On a daily scale, temporal characteristics of instantaneous disasters are much more straightforward, exhibiting clear-cut spikes, compared to a more gradual and non-monotonous trajectory of a red tide. This confirms the results seen on Figure
 \ref{fig: Figure3}, with Twitter activity serving as a much
more consistent proxy for longer-term (e.g.  weekly,  three-day) cumulative red tide conditions as opposed to finer time frequencies (e.g.  daily), although still demonstrating respectable daily correlations at county level.  Secondly, the initial sighting of red tide in June - despite demonstrating sizeable dead fish and respiratory irritation levels in Sarasota area - was accompanied by only a slight increase in relevant tweets (see Figure \ref{fig: Figure2}), which
is something one would rarely expect from an onset of disasters like hurricanes, floods or earthquakes. This likely goes to show the difference between catastrophes in terms of damage severity specifically to the human population, where red tides, while potentially posing huge threats to certain businesses and carrying several negative health impacts, would not be as devastating as the latter, which could lead to extreme property damage and numerous human casualties.

Another interesting observation was that of tweet counts dissipating at faster rates over time compared to the red tide intensity, which is evident on Figure \ref{fig: Figure2}, especially after the third
peak in September. This may have to do with a psychological phenomenon of ”compassion fatigue” (\citeauthor{jacobson2006compassion}, \citeyear{jacobson2006compassion}, \citeauthor{adams2008compassion}, \citeyear{adams2008compassion}, \citeauthor{nakagawa2019pitfall}, \citeyear{nakagawa2019pitfall}), pointing to people, especially younger adults, losing interest in the disaster event if a considerable amount of time passes since its inception, regardless of whether the event is still persisting by that point.  Because of longer duration compared to catastrophes like hurricanes,  earthquakes, floods, 
compassion fatigue could have an even
stronger effect on lowered Twitter activity several weeks after the first sighting of red tide in the area.  With that said, this psychological phenomena is most applicable to people not affected by the disaster directly, e.g.  living further away from the impacted areas.  As opposed to the nationwide scale of the study done in
 \citeauthor{kryvasheyeu2016rapid}, \citeyear{kryvasheyeu2016rapid}, our research focused specifically on the directly impacted locations, hence likely providing a balancing counter-effect to that of compassion fatigue.  All of that would require a specific modeling
adjustment when considering various Twitter metrics and their reliability as a proxy for the actual local red tide conditions, and is intended to be a subject of future research.

When studying the correlations on a county level, there was a slight deterioration in strengths of correlations between red tide conditions and Twitter activity in those more fine-grained areas. That is only natural, due to some locations that didn’t experience the actual
effects of red tide (e.g. Hillsborough) still actively posting about it either in anticipation of the potential impact, or simply reporting on the developments in the neighboring counties. Despite that, utilizing weekly counts of tweets explicitly geo-tagged in the area served as
a great barometer of local beach conditions, showing much less activity in the unaffected areas compared to using all geo-matches, while retaining strong temporal correlations with spikes of red tide in the impacted locations.

In an effort to obtain actionable insight on a more hyper-local scale (e.g. city- or ZCTA-level) and finer temporal scale (e.g.  daily instead of weekly),  we  studied  spatiotemporal correlations of per-capita Twitter activity with local red tide conditions across several localization levels and time frequencies.  An expectedly steady decrease in correlation  strength was observed when going down to more fine-grained locality levels and higher time frequencies, but there was still considerable correspondence for total and county-level data across all temporal scales considered.  City-level correlations were respectable for per-capita explicit geo-tag counts if monitored weekly or every three days.  Posing curious research questions on their own, these aspects may also point to Twitter as a crowd-sourcing tool to augment the already existing practices of place-based monitoring for the actual red tide conditions.

Another critical aspect of using local tweet counts was their strong association with the proximity of tweet location to the areas being impacted by red tide at the time. We demonstrated an exponential decrease in per-capita Twitter activity (on city-level) as the distance to the affected locations increased. Moreover, we also detected a decrease in originality of the posts, meaning the higher percentage of retweets, the further one moved from the affected locations. Looking solely at areas within a 100 miles of disaster impacts (the range of our localized study), our results confirmed the findings from \citeauthor{kryvasheyeu2016rapid}, \citeyear{kryvasheyeu2016rapid} for analogous distances. 

Emotion expressed in the posted tweets might provide us with information that basic tweet counts would not be able to convey. In particular, it could help with better estimation of red tide’s arrival and departure from a particular locale, manifesting in prevalence of
negative and positive messaging originating from the area, respectively.  For those reasons we conducted sentiment analysis and utilized weekly total tweet sentiment as an alternative to weekly tweet counts. When using temporal sentiment as a metric to approximate red tide development on a county level, we witnessed strong (expectedly negative) correlations with the local red tide impacts, especially when considering all geo-matched tweets. Moreover, polarized keyword analysis was indicative of public's high situational awareness, grasping the origins of red tide and severity of the disaster.  With that said, when looking at correlations between tweet sentiments and local beach conditions on multiple spatiotemporal levels, virtually none of them were better than those for tweet count metrics. Considering the extra layer of complexity and uncertainty added by estimating tweet sentiments (see Section \ref{sec:SentAnalysis}) as opposed to simply obtaining counts, it makes the latter a clear choice for most optimal, timely, red tide impact assessment.

Besides sentiment analysis, categorization of tweet content by types of specific concerns being expressed by people was also conducted. Do the locals and tourists mostly cite environmental issues? Or do they primarily refer to economic or health concerns? Ability to answer such questions can in turn have management implications for improving communication about red tide, both general education and real time conditions, and for prioritizing management actions, such as marine debris cleanup, targeted assistance to waterfront businesses and their employees, or distribution of personal protective equipment (PPE). We showed that environmental issues are brought up the most, with health and economic concerns cited considerably less, albeit still constituting popular topics. 

Among threats to validity we should mention that the collected localized Twitter data is not to be fully relied upon as an exact representation of general population living in respective areas. First, it clearly eliminates anyone without a Twitter account, with the reasons ranging all the way from a person being from low-income family without access to technology, to having certain physical or mental disability, to simply a matter of preference to communicate via other social media outlets (Facebook, Instagram, etc). Moreover, in \citeauthor{mislove2011understanding}, \citeyear{mislove2011understanding} it was shown that Twitter tends to significantly underrepresent sparsely populated areas, while also skewing male. These issues have been brought up in the literature before (\citeauthor{morstatter2013sample}, \citeyear{morstatter2013sample}), and while certain papers have worked on approaches to circumvent them (\citeauthor{miranda2015twitter}, \citeyear{miranda2015twitter}, \citeauthor{longley2016geo}, \citeyear{longley2016geo}), it mostly remains a disclaimer warning about limitations of using Twitter data. Lastly, to simplify the data collection and text analysis, in our particular study we focused solely on tweets written in English, which automatically excludes anyone with limited English proficiency or those simply preferring to communicate in a different language.  As studied in \citeauthor{hong2011language}, \citeyear{hong2011language}, Twitter is supremely diverse internationally from linguistic standpoint, and while Twitter activity in Florida is likely to lean heavily English, according to $worldpopulationreview.com$ Spanish is spoken by $21.8\%$ of Floridians, along with Florida having $25.58\%$ of Hispanic population (6th largest in the United States). Hence, our study could be augmented in the future by accounting for tweets sent in Spanish, along with applying Spanish-specific sentiment analysis techniques as discussed in \citeauthor{pla2014sentiment}, \citeyear{pla2014sentiment}).

All-in-all, our analysis of over 18,000 tweets geo-matched to Florida's Gulf coast during the 2017-2019 red tide event showed very strong localized correlations between red tide conditions and public response in terms of the number and sentiment of tweets, along with the percentage of retweets.  While correlations gradually dissipated upon considering more localized scales and higher time frequencies, they were still respectable for spatiotemporal levels that would allow for a potentially timely response (e.g.  county-level on daily basis, or city-level every three days), in general making Twitter a solid proxy of red tide impacts in the area. Provided how reliable the explicitly geo-tagged tweets turned out to be for that purpose, and given their relative ease of data collection (compared to user geoprofile matches, or some of the physical disaster impact measurements), local governments and planning agencies could integrate Twitter into their operations for a relatively low fee and in a rather seamless fashion during disaster events. Creating software pipelines to collect relevant tweets on a daily basis (using SQL \citep{MySQL} and any scripting language), while also calculating counts, evaluating sentiments, categorizing content by concerns being expressed (using R \citep{RCoreTeam}, Python \citep{Python}), creating useful visualizations (using Tableau \citep{TableauPublic}, for example) is a feasible task to be carried out for informing response to the disaster as it unfolds. 

For future work, we plan on conducting further analysis into different Twitter user groups and individual accounts, to characterize social influencers and types of messaging during red tide, in an attempt to identify strategies for improved public communication. Which account types - media, political figures, government bodies, regular citizens - and what kind of messages generate the most activity on Twitter (via ‘likes’, retweets, replies)? Also, how is the messaging and public’s Twitter reaction affected by any other major events happening concurrently with the red tide period (e.g. political elections, other environmental disasters, etc)?  It will be falling more in line with the research work done on public’s perception of red tide
(\citeauthor{kuhar2009public}, \citeyear{kuhar2009public}, \citeauthor{nierenberg2010florida}, \citeyear{nierenberg2010florida}, \citeauthor{kirkpatrick2014florida}, \citeyear{kirkpatrick2014florida}) and social dynamics throughout the event (\citeauthor{mascareno2020twitter}, \citeyear{mascareno2020twitter}).  Moreover, as an alternative to relying solely on geo-tag and user geoprofile information, we plan on utilizing tweets matched to a particular location via explicit mention of that locale in the tweet content. This might potentially open doors for an even better local impact assessment,  along with the ability to gauge the level of interest and study differences in messaging about the event across the nation. Lastly, when conducting sentiment analysis, traditionally one treats tweet replies similarly to the original posts or retweets - as a standalone message. In the future, we could work on incorporating the contents of the original tweet to provide richer context when evaluating sentiment expressed in the reply message.

\section*{SUPPLEMENTARY MATERIALS}

$Supplement.pdf$ file contains supplementary materials referenced in the main manuscript.

To promote additional research on social media applications to inform management response to HAB events, as supplement to this article we provide the complete library of tweets being analyzed in this study, the custom lexicon used for sentiment analysis, and vocabulary of terms for concern types (environment, economy, health, government).

Additionally, a more extensive set of data and source code can be found at \url{https://github.com/UsDAnDreS/Using-Localized-Twitter-Activity-for-Red-Tide-Impact-Assessment.}

\section*{ACKNOWLEDGEMENTS}

This project was supported in part through the Tampa Bay Environmental Restoration Fund (TBERF) and a USEPA Region 4 Cooperative Agreement Grant to the Tampa Bay Estuary Program (CE-00D89319-0). TBEP funding for this project stems from USEPA Section 320 Grant Funds, and the TBEP’s local government partners (Hillsborough, Manatee, Pasco, and Pinellas Counties; the Cities of Clearwater, St. Petersburg, and Tampa; Tampa Bay Water; and the Southwest Florida Water Management District) through contributions to the TBEP’s operating budget. The Sunrise Rotary Foundation of Sarasota also supported this project through a grant to the Science and Environment Council of Southwest Florida. The authors are grateful to Dr. Tracy Fanara for providing the tabular beach conditions data as reported on \url{visitbeaches.org},  Mote Marine Laboratory’s Beach Conditions Reporting  System \citep{LocalBeachData}.

\newpage

\section{APPENDIX}

\subsection{Tweet totals vs per-capita.}



\begin{table}[H]
    \centering
    \begin{tabular}{lll}
\diagbox[width=15em]{Red Tide \\ Condition}{Counts of} & Explicit Geo-Tags & All Geo-Matches\\
\hline
  \begin{tabular}{l}
   \\
       {\it K. Brevis} Cell Counts \\
       Dead Fish Levels \\
       Respiratory Irritation Levels
  \end{tabular} & 
  
  \begin{tabular}{ll}
  Total & Per-Capita\\
  \hline
        0.47 &  0.50 \\ 
        0.73 & 0.79  \\
        0.63 & 0.72 

  \end{tabular} & 
  
  \begin{tabular}{ll}
  Total  & Per-Capita  \\
  \hline
      0.43 & 0.65   \\
      0.47 & 0.72  \\
        0.39 & 0.66 
  \end{tabular}

\end{tabular}

\caption{\textbf{Correlation of weekly red tide conditions with weekly counts of explicitly geo-tagged and all geo-matched tweets, total and per-capita, on the local county-level}. For Tampa Bay surrounding area, including the following counties: Pasco, Hillsborough, Pinellas, Manatee, Sarasota.}
    \label{tab:TabSupp1}
\end{table}


\begin{table}[H]
    \centering
    \begin{tabular}{lll}
\diagbox[width=15em]{Red Tide \\ Condition}{Counts of} & Explicit Geo-Tags & All Geo-Matches\\
\hline
  \begin{tabular}{l}
   \\
       \it{K. Brevis} Cell Counts \\
       Dead Fish Levels \\
       Respiratory Irritation Levels
  \end{tabular} & 
  
  \begin{tabular}{ll}
  Total & Per-Capita\\
  \hline
       0.23 & 0.25  \\ 
       0.58  & 0.63 \\
        0.50  & 0.57 

  \end{tabular} & 
  
  \begin{tabular}{ll}
  Total  & Per-Capita  \\
  \hline
       0.22  & 0.34 \\
      0.39 & 0.60 \\
       0.32   & 0.55 
  \end{tabular}

\end{tabular}

\caption{\textbf{Correlation of daily red tide conditions with daily counts of explicitly geo-tagged and all geo-matched tweets, total and per-capita, on the local county-level}. For Tampa Bay surrounding area, including the following counties: Pasco, Hillsborough, Pinellas, Manatee, Sarasota.
    }
    \label{tab:TabSupp2}
\end{table}


\begin{table}[H]
    \centering
    \begin{tabular}{lll}
\diagbox[width=15em]{Red Tide \\ Condition}{Counts of} & Explicit Geo-Tags & All Geo-Matches\\
\hline
  \begin{tabular}{l}
   \\
       Dead Fish Levels \\
       Respiratory Irritation Levels
  \end{tabular} & 
  
  \begin{tabular}{ll}
  Total & Per-Capita\\
  \hline
      0.39    & 0.54 \\ 
       0.35  &  0.48

  \end{tabular} & 
  
  \begin{tabular}{ll}
  Total  & Per-Capita  \\
  \hline
     0.23  & 0.47  \\
         0.20 & 0.43 
  \end{tabular}

\end{tabular}

\caption{\textbf{Correlation of weekly red tide conditions with weekly counts of explicitly geo-tagged and all geo-matched tweets, total and per-capita, on the local city-level}. We include 57 cities in the Tampa Bay surrounding area, including the following counties: Pasco, Hillsborough, Pinellas, Manatee, Sarasota.
}
    \label{tab:TabSupp3}
\end{table}


\subsection{Multi-Level spatiotemporal correlations of local Twitter activity and with local respiratory irritation levels and K. brevis cell counts.}

\begin{figure}[H]
    \centering
\begin{minipage}{0.47\textwidth}
\textbf{A}

\includegraphics[scale=0.65]{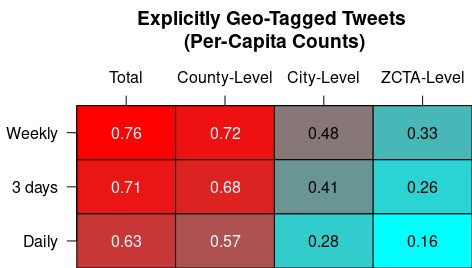} 

\

\

\

\textbf{B}

\includegraphics[scale=0.65]{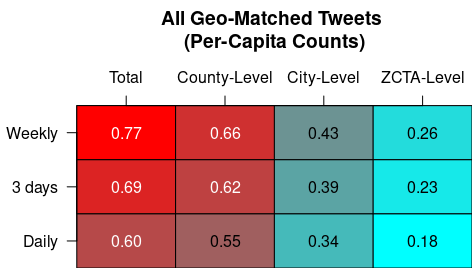} 

\end{minipage} \ \ \ \ \ \ \ 
\begin{minipage}{0.47\textwidth}
\textbf{C}

\includegraphics[scale=0.65]{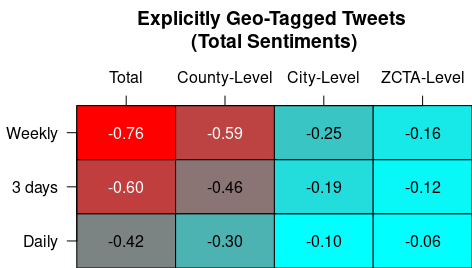} 

\

\

\

\textbf{D}

\includegraphics[scale=0.65]{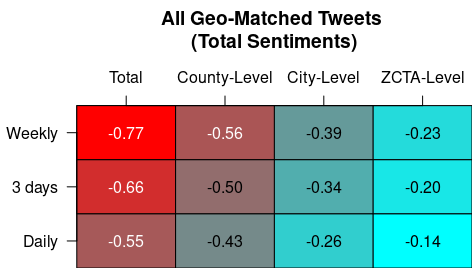} 
\end{minipage}

\

\

     \caption{\textbf{Multi-level spatiotemporal correlations of local Twitter activity with respiratory irritation levels observed in the area.} \textbf{A}: Heatmap-style matrix of spatiotemporal correlations for per-capita counts of tweets explicitly geo-tagged in the area with local red tide conditions, calculated on different levels of locality and time frequency. \textbf{B}: Same as A, but for per-capita counts of all tweets geo-matched to the area, hence also including those matched by user's geoprofile information. \textbf{C}: same as A, but for per-capita total tweet sentiments of explicitly geo-tagged tweets. \textbf{D}: same as B, but for per-capita total tweet sentiments of all tweets geo-matched to the area.
     }
    \label{fig:FigureSupp1}
\end{figure}

\begin{figure}[H]
\centering
\begin{minipage}{0.47\textwidth}
\textbf{A}

\

\includegraphics[scale=0.22]{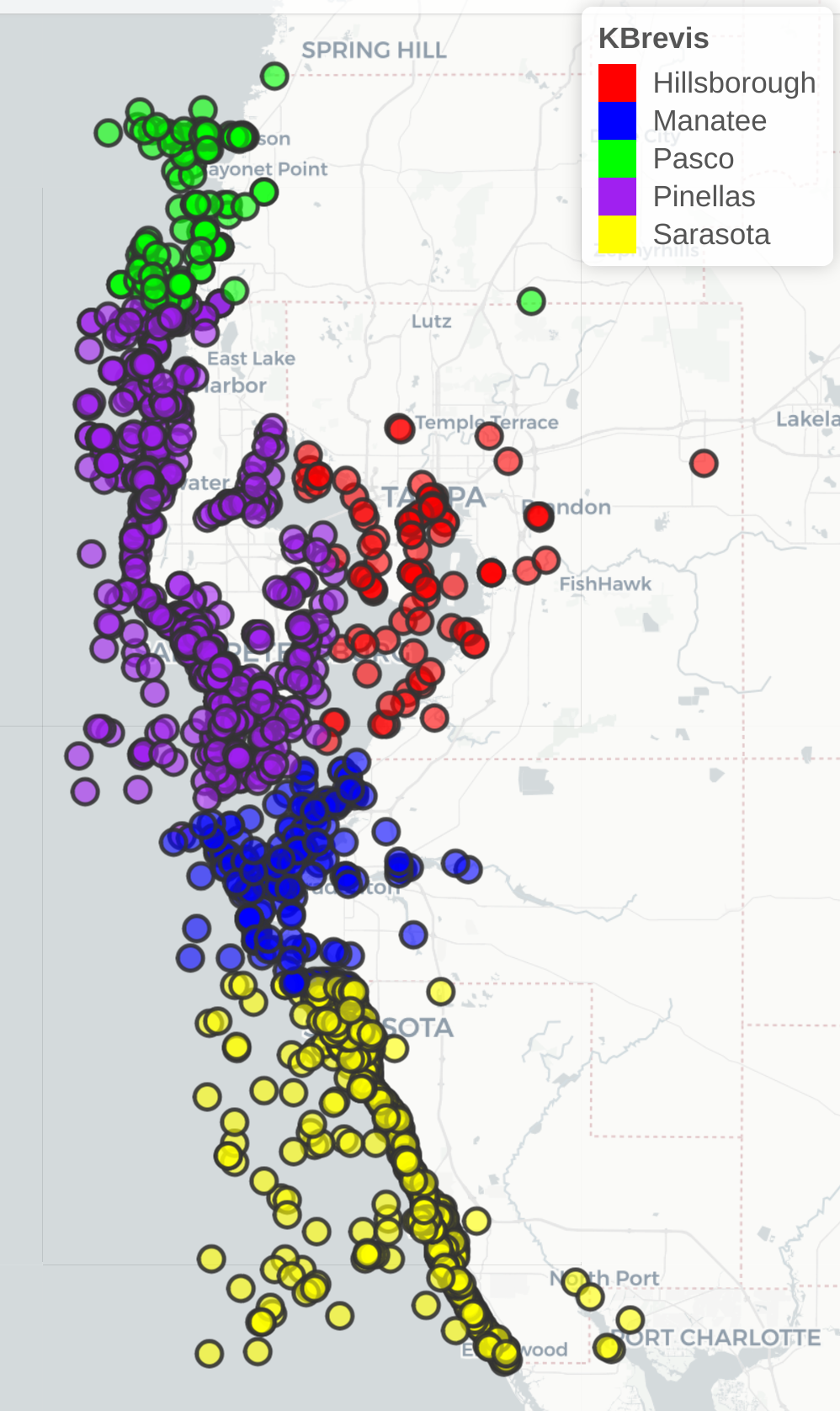}

\end{minipage} \  \ \
\begin{minipage}{0.24\textwidth}

\textbf{B}

\

\includegraphics[scale=0.55]{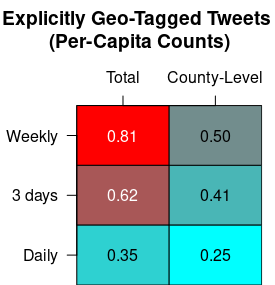}

\

\

\textbf{C}

\

\includegraphics[scale=0.55]{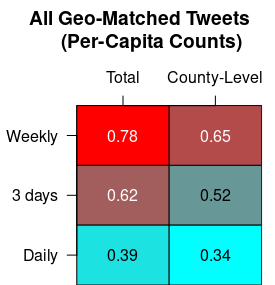}

\end{minipage}   \
\begin{minipage}{0.24\textwidth}

\textbf{D}

\

\includegraphics[scale=0.55]{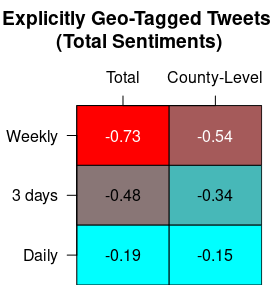}

\

\

\textbf{E}

\

\includegraphics[scale=0.55]{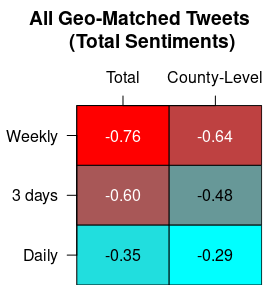}

\end{minipage}
   \caption{\textbf{Multi-level spatiotemporal correlations of local Twitter activity with K. brevis levels observed in the area.} \textbf{A}: Map of {\it K. brevis} sample locations, with their county assignments. \textbf{B-C}:  Heatmap-style matrices of spatiotemporal correlations for per-capita counts of explicitly geo-tagged tweets (B) and all tweets geo-matched to that area (C). \textbf{D-E}: Heatmap-style matrices of spatiotemporal correlations for total sentiments of explicitly geo-tagged tweets (D) and all tweets geo-matched to that area (E).
   }
    \label{fig:FigureSupp2}
\end{figure}

\subsection{Example of daily red tide dynamics}
\label{sec:DailyRedTideDynamics}

\begin{figure}[H]
    \centering
    \includegraphics[scale=0.65]{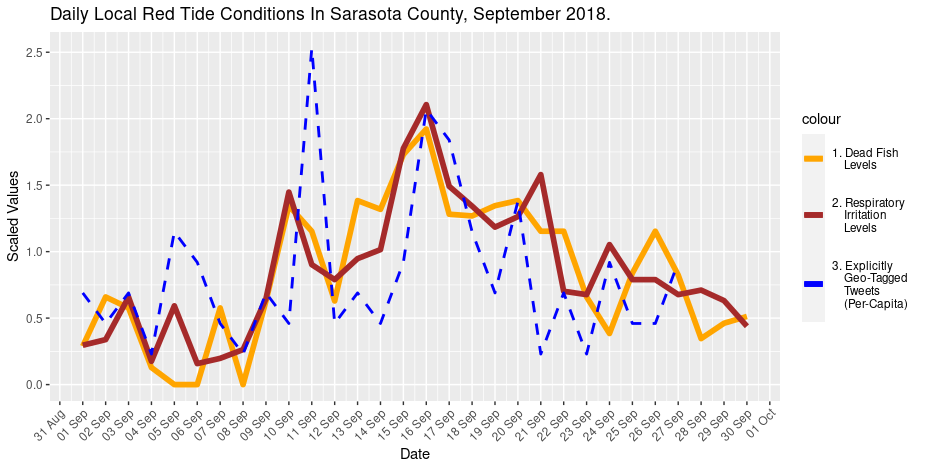}

    \caption{\textbf{Daily local red tide conditions and tweet counts in Sarasota during September 2018.} Local {\it K. brevis} cell count data was not available for every single day, hence was excluded from the graphic.}
     \label{fig:DailyTemporalDynamics}

\end{figure}

\bibliographystyle{model2-names}
\bibliography{bibliography.bib}

\end{document}


\def\spacingset#1{\renewcommand{\baselinestretch}%
{#1}\small\normalsize} \spacingset{1}

\date{}

  \title{\bf 
  Supplementary Materials for "Using Localized Twitter Activity for Red Tide Impact Assessment"
  }
    \maketitle
\newpage

\date{}

\section{Red tide tweets within the context of 2018 Florida elections}
\label{sec:PoliticalContext}

\begin{figure}[]
    \centering
    \includegraphics[scale=0.21]{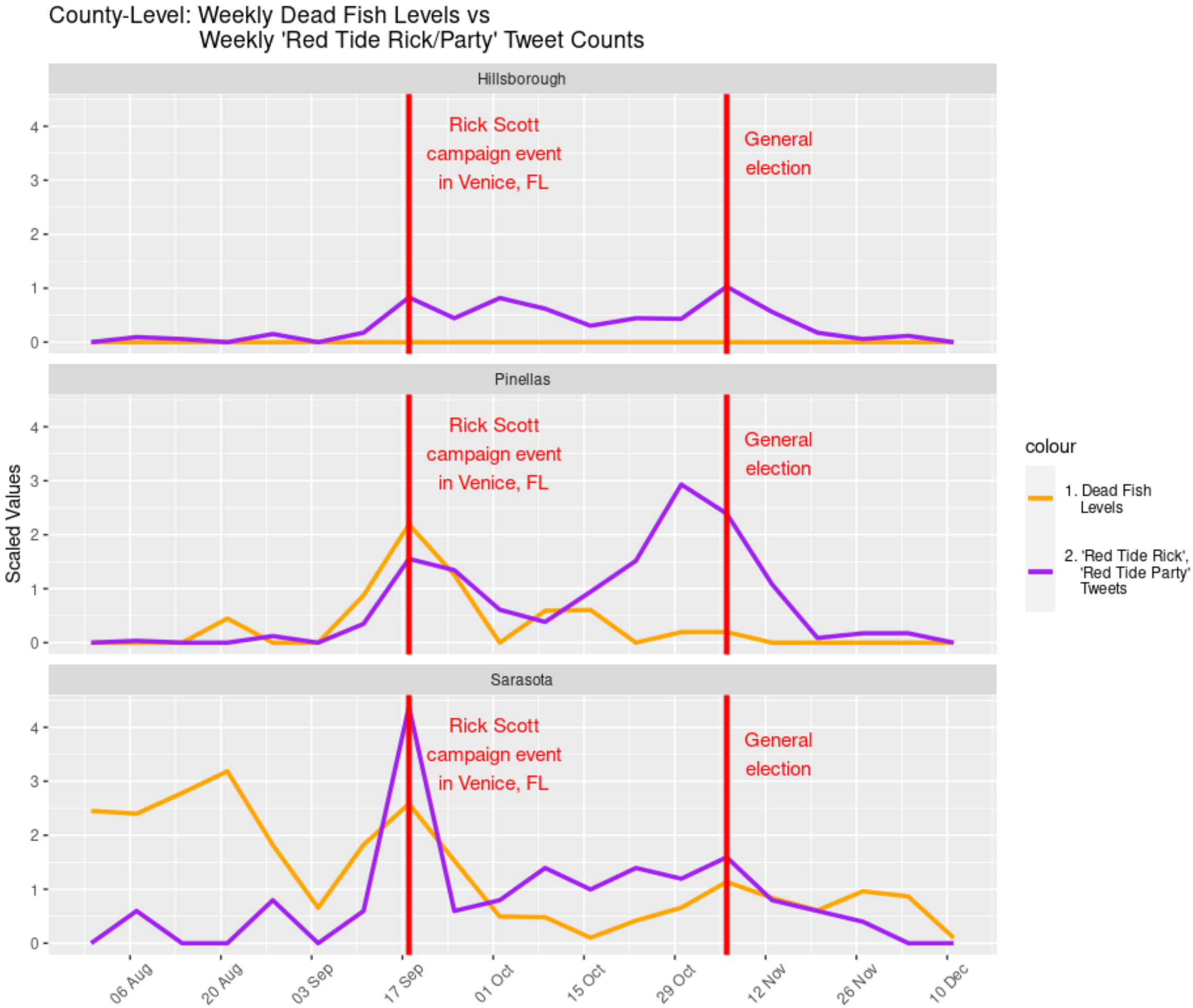}
\caption{\textbf{County-level weekly correspondence of "Red Tide Rick" and "Red Tide Party" tweets with dead fish levels and select political events.}}
    \label{fig:RedTideRickParty_Tweets}
\end{figure}

The Florida 2018 red tide event overlapped with the state-wide elections, which could considerably impact the messaging of Twitter users. We discussed how tweets having political nicknames such as "Red Tide Rick" or "Red Tide Party" as their sole reference to red tide would be more indicative of local political events rather than the red tide conditions, and Figure \ref{fig:RedTideRickParty_Tweets} demonstrates precisely that for the counties where these mentions were most prevalent. Peaks of "Red Tide Rick"/"Red Tide Party" term usage were observed to be extremely close to either the general election (Nov 6th), including the lead-up to it, and Rick Scott's US Senate campaign event in Venice, FL (Sep 18th), which was accompanied by protests from local residents (with tweet contents clearly reflecting these events). While some of these spikes of activity also correspond to increases in dead fish levels (e.g. in the case of Sarasota county), mostly we see clear examples of such tweets being dictated by political events rather than evidence of worsened red tide conditions (see Hillsborough, or Pinellas in the lead-up to general election). Hence, we ended up excluding such tweets.

\begin{figure}[]
    \centering
    \includegraphics[scale=0.7]{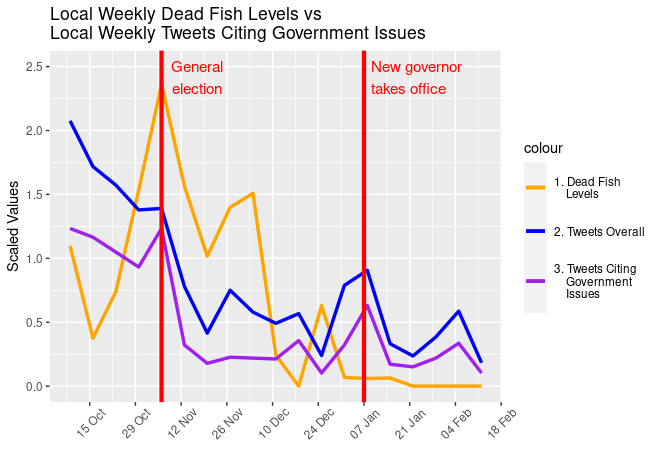}
\caption{\textbf{Weekly correspondence of tweets (overall, and those citing government concerns) with dead fish levels and select political events, for the entire Tampa Bay surrounding area.}}
    \label{fig:Political_Tweets}
\end{figure}

Additionally, to further illustrate the impact that 2018 local elections had on Twitter messaging, on Figure \ref{fig:Political_Tweets} we demonstrate how tweets citing government issues drive the overall Twitter activity on dates of political events. Specifically, in the lead-up to the general election, while the overall load of tweets relevant to red tide is going down, messages mentioning government and elections don't dissipate and peak during the election week. Meanwhile, the week of January 7th, we witness a clear spike in government-related tweets, which matches with the new Florida governor taking office. These tweets emphasize red tide as one of the primary issues for the new governor to combat, and mention him signing the executive order to fight red tide and toxic algae. While relevant to the big picture of the red tide as an issue moving forward, these tweets don't necessarily reflect local impacts at that particular point in time, and hence might need certain modeling adjustments as a topic of future research.

\section{Things that didn't work: user account types, lead/lag correlations.}

Among several other things we have tried that didn't come to fruition were the considerations to focus on correlations of red tide conditions with tweets from specific account types (e.g. only media, only citizens), and
studying the delayed reactions and anticipation of red tide impacts via lag and lead correlations,
respectively (\citeauthor{spriggs1982lead}, \citeyear{spriggs1982lead}, \citeauthor{dajcman2013interdependence}, \citeyear{dajcman2013interdependence}). For the former, we hypothesized that tweets from citizen accounts should provide more localized reporting of conditions, while the media is more incentivized to cover the red tide proceedings across the entire coast (regardless of where the media members themselves are located). Therefore, we divided all Twitter accounts into two main categories: media and citizens. There were other account types that didn't fit into either user group, e.g. political figures, or organizational and government accounts, but they were too few for analysis to be deemed trustworthy. Due to counts of explicitly geo-tagged tweets having proved to be the most
robust metric, we utilize it for our user group comparisons in Table \ref{tab:Table2}. Neither citizen- nor media-only tweet counts consistently beat the simpler approach of just including all the
tweets in terms of county-level correlations with dead fish levels. In fact, even beyond the cases
considered in the table below (correlations of explicitly geo-tagged tweets with dead fish levels,
broken down by county), all tweets come out on top, or at the very least not worse, compared
to citizen- and media-only tweets. Using tweet sentiments as the metric produced similar results.


\begin{table}[h]
    \centering
    \begin{tabular}{llll}
\diagbox[width=14em]{Time Frequency}{Account Type} & Everyone & Citizens & Media\\
\hline
   Weekly  & 0.79  & 0.80  & 0.70 \\
  3 days &  0.74  & 0.71  & 0.67\\
  Daily & 0.63 & 0.59 & 0.53\\
 
\end{tabular}

    \caption{\textbf{County-level correlation of weekly, three-day, and daily dead fish levels with the respective counts for tweets explicitly geo-tagged in the Tampa Bay surrounding area, by account type}.
    }
    \label{tab:Table2}
\end{table}


Finally, given the time series nature of our data, and the importance of being able to both
anticipate and retrospectively evaluate red tide effects, we were also interested in calculating
lag and lead correlations between Twitter metrics and local conditions.  In other words, we wanted to know if online user activity during current period (be it a day, three days, or a week) could be used to better gauge the impacts taking place in the previous period (lagged reaction), or to anticipate red tide appearance in the area in the near future. This analysis provides insight as to how quickly the information travels to Twitter, with higher lagged correlations indicating a delayed social media reaction to the actual red tide impacts. In Table \ref{tab:Table3}
you see the correlations with local dead fish levels of three-day per-capita tweet counts for all explicitly geo-tagged tweets, across several localization levels. One may witness these lead and lag correlations being consistently worse than the concurrent correlations, pointing to lack of evidence for especially strong delayed reaction or anticipatory quality, respectively, of Twitter activity. While Twitter activity could be potentially used to anticipate the impacts, simple correlation is not the preferred forecasting tool in that regard (one might want to work with the proper time series forecasting models as per \citeauthor{chatfield2000time}, \citeyear{chatfield2000time}, and \citeauthor{de200625}, \citeyear{de200625}).


\begin{table}[H]
    \centering
    \begin{tabular}{llll}
\begin{tabular}{c}
     Dead Fish Levels  \\
     Over 
\end{tabular} &

\begin{tabular}{l}
   Current 3 Days    \\
    (Concurrent)  
\end{tabular}  & \begin{tabular}{l}
   Last 3 Days    \\
    (Lag)  
\end{tabular} & \begin{tabular}{l}
   Next 3 Days    \\
    (Lead)  
\end{tabular}\\
\hline
   Total  & \ \ 0.78   & \ \ 0.74   & \ \ 0.73 \\
   County-Level  & \ \ 0.74  & \ \ 0.68   & \ \ 0.70\\
   City-Level & \ \ 0.46  & \ \  0.42  & \ \ 0.42\\
  ZCTA-Level & \ \ 0.28 & \ \ 0.25 & \ \ 0.26\\
\end{tabular}
    \caption{\textbf{Correlation of counts for tweets explicitly geo-tagged in the Tampa Bay surrounding area over the current three days with the dead fish levels over the current (concurrent), previous (lag) and next (lead) three days.}}
    \label{tab:Table3}
\end{table}


\bibliographystyle{model2-names}
\bibliography{bibliography.bib}